\documentclass{aa}
\usepackage{graphics}
\usepackage{amssymb}
\usepackage{natbib}    
\usepackage{journals}
\newdimen\digitwidth
\setbox0=\hbox{\rm0}
\digitwidth=\wd0
\catcode`#=\active
\def#{\kern\digitwidth}
\begin{document}
\thesaurus{03(11.03.1; 11.03.4 HCG~16; 13.25.2; 13.25.3)}
\title{Clumpy diffuse X-ray emission from the spiral-rich compact galaxy
group HCG~16}
\author{S. Dos Santos\inst{1} \and G. A. Mamon\inst{1,2}}
\institute{Institut d'Astrophysique (CNRS UPR 341), 98 bis Bd Arago,
F--75014 Paris, FRANCE  ({\tt santos,gam@iap.fr})
\and DAEC (CNRS UMR 8631), Observatoire de Paris, F--92195 Meudon, FRANCE}
\offprints{S. Dos Santos}
\date{Received ...... / Accepted ......}
\titlerunning{Clumpy diffuse X-ray emission from HCG~16}

\maketitle
\begin{abstract}

We carefully 
reanalyze the ROSAT PSPC X-ray spec\-tro-photo\-metric observations of HCG
16 (Arp 318), and compare them to optical and radio data.  
Its X-ray morphology resembles its morphology at 20$\,$cm, seen by the NVSS.
In particular,
we detect diffuse emission in eight regions filling half of the
$200\, h_{50}^{-1} \,{\rm kpc} \ (8\farcm 7)$ radius circle around the optical
center of the group: one region 
encompassing galaxies a \& b, two regions 
surrounding the group galaxies c \& d, a clumpy region
roughly $140\,h_{50}^{-1}\,\rm kpc$ from the group galaxies, which
may be gas ejected from one of the galaxies, plus regions respectively
associated with a background 
radio-source, a probable background radio-source, a foreground star and a
background group or cluster.
The
bolometric X-ray luminosity of the diffuse emission, excluding the regions
associated with radio galaxies, is $L_X^{\rm bol} = 2.3 \times
10^{41}\,h_{50}^{-2} \rm
erg\,s^{-1}$, {\it i.e.\/,} half of the
luminosity found by \cite{PBEB96}.  
The region that is offset from the galaxies
contributes half of the diffuse X-ray luminosity of the group.  The
diffuse emission is cool ($T < 0.55\,\rm keV$ with 90\% confidence with a
best fit $T = 0.27\,\rm keV$).
At these low temperatures, the correction for photoelectric
absorption in the estimate of
bolometric luminosity is a factor 3.5 and varies rapidly with temperature,
hence an uncertain bolometric luminosity.
%

The clumpy distribution of hot diffuse gas in HCG~16 is illustrated by the low
mean X-ray surface brightness and hot gas density of the
regions of undetected emission within $8'$ (at most 1/4 and 1/6 of those
of the detected gas, assuming both have same temperature, metallicity and
clumpiness).
The irregular 
X-ray morphology of the diffuse emission rules out a (nearly) virialized
nature for HCG~16, unless intergalactic gas  had sufficiently high
specific entropy to be unable to collapse with the group.
In any event, the clumpy gas distribution, and high luminosity given the low
temperature suggest that most of the diffuse gas originates from galaxies,
either through tidal stripping or through galactic winds driven by supernova
remnants. 
Therefore, no spiral-only HCGs are known with regular diffuse
emission tracing a gravitational potential.

Our results highlight the need for a careful 2D spatial analysis and
multi-wave\-length study of the
diffuse X-ray emission from groups, suggesting that other compact
groups could be significantly contaminated by superimposed X-ray sources.

\end{abstract}


\section{Introduction}
\label{introd}

The extreme apparent density of
compact groups of galaxies makes them ideal sites for the study of multiple
interactions of galaxies, and more generally, another dense environment to be
compared with rich clusters. Indeed, compact groups (of typically 4--5 bright
galaxies) such as those (hereafter, HCGs) 
cataloged by \cite{H82} appear as compact in projection onto the plane of
the sky as the cores of rich clusters, and are moreover selected to be
isolated. 


The discovery of numerous signs of galaxy-galaxy interaction within 
HCGs (see 
\citealp{H97rev} for a review) suggest that most HCGs are
indeed dense in 3D. However, the very short crossing times
derived from galaxy spectroscopy \citep{HMHP92} suggest rapid galaxy
merging and coalescence into a single giant
elliptical galaxy \citep{CCS81,Barnes85,M87,Barnes89,BBCL94,GTC96}
if indeed they correspond to the low-mass end of clusters of
galaxies, forming at high redshift,
unless the group is constantly replenished through accretion of surrounding
galaxies
\citep{DGR94,GTC96}. 
In fact, a variety of arguments have been 
put forth suggesting that 
compact groups are mostly chance
alignments of galaxies along the line of sight within larger systems: loose
groups 
(\citealp{Rose77} for chain-like groups;
\citealp{M86} for the majority of compact groups), 
clusters \citep{WM89} or cosmological 
filaments \citep*{HKW95}.
Thanks to gravity, these chance alignments tend themselves to be binary-rich
\citep{M92_DAEC}, and it is very difficult to tell whether the interaction seen
in HCGs are caused by a system of 4 or more bright
galaxies or simply by binaries, well-separated along the line-of-sight.

Recently, there has been much hope that the debate on the nature of HCGs
could be 
resolved by X-ray observations.  The ROSAT, ASCA and Beppo-SAX X-ray
satellites are 
sensitive enough in the soft X-ray 
band to be able to detect the diffuse intergalactic plasma within
(nearly) virialized galaxy systems with potential wells with depth
corresponding to a 1D velocity dispersion of $\sim 250 \, \rm km \, s^{-1}$.
Most sensitive of these is the {\sf Position Sensitive Proportional Counter}
({\sf PSPC}) of ROSAT, which discovered diffuse intergalactic
emission from large numbers of compact groups
(\citealt*{EVB94}; \citealt*{PBE95}; \citealt*{SC95}; \citealt{MDMB96};
\citealt{PBEB96}, hereafter PBEB).

But there has been a debate on whether the fraction of compact groups
detected by ROSAT is
40\% \citep{MDMB96} or 75\% (PBEB, taking into account the selection effects on
distance).  There is also controversy on the luminosity-temperature relation:
\cite{MZ98} 
derive a relation that is consistent with the
extrapolation from rich clusters, while PBEB find a luminosity-temperature
relation for compact groups with a much higher slope, with the hottest groups
lying on the cluster extrapolation.


Moreover, the morphology of the diffuse X-ray emission of compact groups 
is very diverse, as attested by the {\sf PSPC}
maps provided by \cite{PBE95}, \cite{SC95}, \cite{PB93} (only for HCG62) and
PBEB (only for 
HCG~16).  In some cases, such as HCG 62 
(\citeauthor{PB93}), diffuse intergalactic emission extends well
beyond the group, not centered on any galaxy of the group, with regular
circular isophotes, just as is to be expected in a well relaxed galaxy
system.  But in other cases, the emission is only attached to individual
galaxies, see {\it e.g.} HCG~44 (PBEB).  
And there are intermediate cases, such as
HCG~16,  
where the diffuse emission does 
not appear as extended as the galaxy system nor as regular as in HCG 62.

In fact, HCG~16 (also known as Arp 318) is an unusual galaxy
system.
First, because the six brightest galaxies of the group (Hickson's original 4
plus two more outside the group isolation annulus, see \citealp{dCRZ94}) 
are starburst, LINERs or
AGNs \citep[][hereafter RdC3Z]{RdC3Z96}.
Moreover, the X-ray properties of HCG~16 are controversial and possibly
extreme.
It was 
first detected with the EINSTEIN satellite \citep{BHR84},
which did not have the angular resolution to resolve the emission between the
group galaxies and an intergalactic medium.
However, \citeauthor{PBEB96}'s analysis of ROSAT {\sf PSPC} observations made
it  
the coldest detected group ($T = 0.30\pm 0.05\,\rm
keV$), and there are no other spiral-only compact groups with diffuse X-ray
emission 
(\citealp{Mulchaey99}; see also \citeauthor{PBEB96}).
HCG~16 is thus an abnormal group given 
the very strong correlations between X-ray luminosity and group spiral
fraction found by \cite{PBE95} and \cite{MDMB96}.
Moreover, whereas diffuse X-rays were clearly
detected by PBEB, \cite{SC95} failed
to detect such diffuse emission at an upper 
limit 16 times 
lower,\footnote{Given the fluxes measured by \cite{SC95}
for HCG~16 and their adopted value for $H_0$, their quoted upper 
limit for their
luminosities were underestimated by a factor 2 for all undetected
groups in their Table~4 except HCG 3.}
whereas only a factor 2.3 (which we find by simulating a {\sf MEKAL} plasma
with temperature, abundance and absorbing column as quoted by PBEB)
is attributable to the wider (``bolometric'') energy range
in which
PBEB compute their luminosities.
Given the low temperature that PBEB derive for HCG~16,
their derived X-ray luminosity places it two orders of
magnitude above their compact group
luminosity-temperature relation and roughly a factor of two above  the
extrapolation of the cluster trend.
It thus seems difficult to reconcile HCG~16 with a low temperature
extrapolation of regular X-ray emitting compact groups.
Indeed, PBEB note that HCG~16 is ``probably not fully virialized''.

In this article, we present a detailed analysis of the ROSAT {\sf PSPC}
observations of HCG~16.
The data reduction is presented in Sect.~2, our spatial analysis in Sect.~3,
and our spectral analysis in Sect.~4.
In Sect.~5, we compare our results with previous X-ray analyses of HCG~16,
perform a mass budget of the group and ask if it is virialized.
In a following paper \citep[][hereafter Paper~II]{MDS99}, 
we discuss at length the cosmological and dynamical constraints on the
nature of HCG~16.

\section{Observations and data reduction}

\subsection{Observations and preliminary reduction}

\label{datared}
HCG~16 was observed in January 1992 with the {\sf PSPC} (in its low-gain
state), on 
board the ROSAT 
satellite, for a total observing time of $14\,634 \,{\rm s}$. 
We obtained the data from the ROSAT archives. 
\cite{SMBM94}'s {\sf PSPC Extended Source Cookbook}
software
was used to perform the first-pass data reduction, {\it i.e.\/,} 
rejection of high-background times, energy-depen\-dent (in 7 bands)
background subtraction,
exposure and vignetting corrections.
We adopted a
conservative value of $170 \, {\rm cts\,s^{-1}}$ for the maximum Average
Master Veto rate  allowed  \citep[see][]{SMBM94}. Even with
this low threshold, 
only   
$6\%$ of the total observing time was rejected, leaving an effective
observational time of $13 \, 748 \, {\rm s}$.
We then carefully examined the light curves of the total counts in the entire
image per
energy band as defined by \cite{SMBM94}, and checked that no short
time scale glitches were present.
Point sources
were detected using a sliding box algorithm, with the improvement that the
box is a circle with a radius varying with off-axis angle, to model the
varying point spread function. Each point source detected at a level exceeding
$3 \, \sigma$ was removed, {\it i.e.} a circle centered on the source, of
radius $1.5$ times the $90 \, \%$ encircled energy radius, was excised.
Unless otherwise stated, we limited our analysis to the $[0.2,2.0] \,\rm keV$
energy band, because 
at lower energies, the background is too high to be adequately subtracted
from the data, while at high energies the same occurs because our sources
happen to be relatively cold, and moreover the calibration of the {\sf PSPC}
is uncertain.
This preliminary reduction produced an image with $512 \times 512$ pixels,
$15 \arcsec$ wide (roughly the FWHM of the {\sf PSPC}'s PSF 
at $1 {\rm keV}$).

\subsection{Background estimation}
\label{bgest}

The definition of the background region, whose counts are subtracted to each
pixel within some region, is of crucial importance for the spatial detection of
sources with very low signal-to-noise ratio (hereafter S/N), as well as for
spectral analyses. 

The shadowing by the supporting ring of the {\sf PSPC} (situated at $\simeq
20'$ 
from the center of the field) is visible in the
images, even after the vignetting correction. Thus, we measure the background
well outside of the ring.
In practice, we choose three annular regions to measure the background, each
centered on HCG~16, with inner and outer radii of $30'-48'$ (BG1), $26'-40'$
(BG2), and $26'-34'$ (BG3).
The radial structures of the {\sf PSPC}
supporting 
ring were removed in each case.
Table~\ref{BGtab} shows the background counts in
the three regions 
within the $0.2-2.0\,\rm keV$ energy range.
We note that BG3 has a slightly lower value than the
other two. 

\begin{table}[ht]
\caption{Background from different annular regions}
\begin{tabular}{lccc}
\hline
Region & Radii & Pixels & Counts/pixel \\  
\hline
BG1 & $30 \arcmin - 48 \arcmin$ & $25930$ & $0.593 \pm 0.005$ \\
BG2 & $26 \arcmin - 40 \arcmin$ & $17630$ & $0.577 \pm 0.006$ \\
BG3 & $26 \arcmin - 34 \arcmin$ & $9832$ & $0.565 \pm 0.008$ \\
\hline
\end{tabular}
\label{BGtab} 
\end{table}

To decide which is the best background, we measure the \emph{net} counts in two
regions within the inner ring support, using each of the three background
regions for measuring the background. The latter
have been vignetting corrected and normalized to the number of pixels of each
region. Our two test regions are the $8 \arcmin$ radius circle 
centered on the group optical center and
the annulus surrounding this circle, with inner and outer radii of $8
\arcmin$ and $17 \arcmin$. Both are in the inner $20 \arcmin$
of the field of view, thus avoiding problems 
with the supporting ring of the {\sf PSPC} and with somewhat uncertain
vignetting correction outside of this ring.
Table~\ref{BGtest} gives the background subtracted data
and errors for both regions and for each background.

\begin{table}[ht]
\caption{Tests of background subtraction}
\tabcolsep 4pt
\begin{tabular}{lccrr}
\hline
Region & \multicolumn{2}{c}{$ r < 8'$} & \multicolumn{2}{c}{$8' < r < 17'$} \\
 & Net counts & Significance & Net counts & Significance \\
\cline{2-3}
\cline {4-5}
\hline
BG1 & $222 \pm 48$ & 4.6$\,\sigma$ & $-250 \pm #78$ & $-3.2\,\sigma\ \ \ \ $ \\
BG2 & $271 \pm 49$ & 5.5$\,\sigma$ & $-121 \pm #82$ & $-1.5\,\sigma\ \ \ \ $ \\
BG3 & $308 \pm 51$ & 6.0$\,\sigma$ & $#-26 \pm #91$  & $-0.3\,\sigma\ \ \ \ $
\\ 
\hline
\end{tabular}
\label{BGtest}

\end{table}

While we find positive net counts within the inner $8'$ of HCG~16, whatever
three 
of the background regions is used to estimate the background, only BG3 is
compatible with the counts within the $8'-17'$ annulus, yielding near zero net
counts. The net counts with the other two background regions are difficult to
understand, unless there happens to be X-ray absorption by Galactic or
intergalactic neutral hydrogen merely in this $8\arcmin - 17 \arcmin$
annulus. This appears to be ruled out by HI observations  with the VLA
\citep{Williams98,VM99} and
by the spectral analysis (Sect.~\ref{spectral}). 
We thus infer that BG1 and BG2 are contaminated by sources or
suffer from uncertain (large) vignetting corrections.
In fact, \cite{PBE95} had also encountered a {\sf PSPC} field with a
background that rose with distance to the field center.
We therefore use BG3 ($26' < r < 34'$) to measure the background, which then
amounts to $B = 6.6\pm 0.1 \times 10^{-4} \,\rm s^{-1}\,arcmin^{-2}$ (in the
$[0.2-2.0]\,\rm keV$ energy range).

\section{Spatial analysis}



\subsection{Preliminary spatial analysis}

\cite{PBEB96} claimed a $3 \,
\sigma$ detection of diffuse emission within a radius $\simeq 8\farcm 5$
surrounding HCG~16 (corresponding to 
$\simeq 195 \, h_{50}^{-1} \, {\rm kpc}$, given HCG~16's redshift of
$0.0132$\footnote{Distances throughout this paper are derived assuming a
Hubble constant $H_0 =
50 \,h_{50} \,{\rm km \, s^{-1} \, Mpc^{-1}}$, with $h_{50} = 1$, unless
explicitly given.}). 
The net counts shown in Table~\ref{BGtest}
confirm PBEB's global detection of diffuse emission in HCG~16.
We detect diffuse emission at $6\,\sigma$ within $8'$ and fail to detect
significant counts between $8'$ and $17'$.
Hence, to first order, the extent of the diffuse emission is very roughly 
$8'$, or $190\, h_{50}^{-1} \,{\rm kpc}$, similar to what is
inferred from PBEB's surface brightness profile of HCG~16.

But we wish to go further: what is the spatial distribution of this excess of
photons within $8 '$ of the group center?
Here we are confronted with the low number of detected photons ($\sim 300$). 
We applied three different methods to
analyze the photon distribution within $8'$.
First, we count the photons
in a grid encompassing the $8 '$ radius circle. Then, in order to
obtain the best S/N ratio, we adaptively smooth the image and detect diffuse
emission. Finally, we apply a wavelet-based method to detect structures at
all scales and verify that emission is present on more than one scale as a
confirmation of its diffuse nature.

\subsection{Analysis on a grid}
\label{grid}

\begin{figure}[bt]
\rotatebox{-180}{\resizebox{!}{\hsize}{\includegraphics{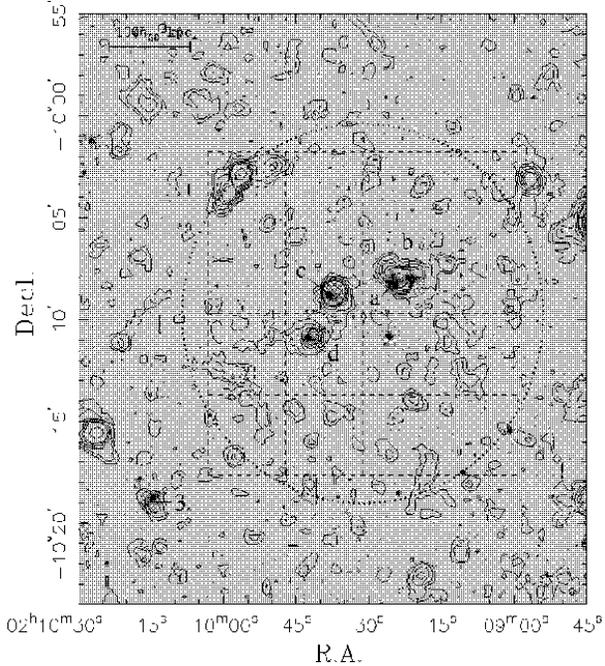}}}
\caption{Contour map of smoothed (by a Gaussian of FWHM = $45 \arcsec$) 
ROSAT-{\sf PSPC} X-ray
emission from HCG~16 superimposed on an optical image of the group provided by
the {\sf Digital Sky Survey} ({\sf DSS}). 
Coordinates are for epoch J2000.
This figure and the following ones do not mask the detected point sources
(which are masked in the subsequent X-ray photometry and spectral analyses).
Contours are drawn at $1 \, \sigma, 2\,
\sigma, 3\, \sigma, 4\, \sigma, 5\, \sigma, 10\, \sigma, 15\, \sigma, 20\,
\sigma, 30\, \sigma, \hbox{ and } 
50\, \sigma$ above the background level, where
$\sigma$ 
is the standard deviation of the smoothed background. The scale (\emph{upper
left}) 
is for the plane at the distance of HCG~16. The four 
original galaxies discovered by Hickson \citeyearpar{H82} are marked `a',
`b', `c'
and `d', together with galaxy HCG~16--3 \citep{dCRZ94}, marked `3', 
which is at the same distance as the group.
The {\it dotted circle\/} centered on the group optical center is
$8 \farcm 7$ wide, {\it i.e.\/,} $200 h_{50}^{-1} {\rm kpc}$ at
the distance of HCG~16.
The {\it dashed square grid\/} used for the preliminary
spatial analysis has cells $16$ pixels wide ($92 \, h_{50}^{-1}
{\rm kpc}$). Cells are counted from left to right then from top to bottom.}
\label{gridfig}
\end{figure}

The simplest way to spatially analyze HCG~16 is to define a grid
encompassing the whole group. We divide the field of view of the group in
sixteen $16\times16$ pixel squares,  ({\it i.e.\/,} $92\, h_{50}^{-1} {\rm
kpc}$ side at the distance of HCG~16).  
The grid overlaid on a smoothed X-ray
image of HCG~16 is displayed in Fig.~\ref{gridfig}. After removing the
point sources (including group galaxies), we counted the
background-subtracted number of photons in each square of the grid. The
results, as well as the statistical significance of the detections, are given
in Table~\ref{gridtab}.  

\begin{table}[ht]
\caption{Net counts within a square grid}
\begin{tabular}{lcrr}
\hline
Region & Pixels & \multicolumn{1}{c}{Net counts} &
\multicolumn{1}{c}{Significance ($\sigma$)}\\ 
\hline
$1$ & $106$ & $-4 \pm #9$ & $-0.5$ 
\ \ \ \ \ \ \ \ \ \  \\
$2$ & $252$ & $-9 \pm 13$ & $-0.7$ 
\ \ \ \ \ \ \ \ \ \  \\
$3$ & $204$ & $-3 \pm 12$ & $-0.2$ 
\ \ \ \ \ \ \ \ \ \  \\
$4$ & $217$ & $17 \pm 13$ & $1.3$ 
\ \ \ \ \ \ \ \ \ \  \\
\hline
$5$ & $256$ & $-8 \pm 13$ & $-0.6$ 
\ \ \ \ \ \ \ \ \ \  \\
$6$ & $198$ & $23 \pm 13$ & $1.8$ 
\ \ \ \ \ \ \ \ \ \  \\
${\bf 7}$ & ${\bf 167}$ & ${\bf 31} \pm {\bf 13}$ & ${\bf 2.5}$ 
\ \ \ \ \ \ \ \ \ \  \\
${\bf 8}$ & ${\bf 256}$ & ${\bf 57} \pm {\bf 16}$ & ${\bf 3.7}$ 
\ \ \ \ \ \ \ \ \ \  \\
\hline
${\bf 9}$ & ${\bf 200}$ & ${\bf 50} \pm {\bf 14}$ & ${\bf 3.6}$ 
\ \ \ \ \ \ \ \ \ \  \\
$10$ & $203$ & $15 \pm 13$ & $1.2$ 
\ \ \ \ \ \ \ \ \ \  \\
${\bf 11}$ & ${\bf 194}$ & ${\bf 34} \pm {\bf 13}$ & ${\bf 2.5}$ 
\ \ \ \ \ \ \ \ \ \  \\
${\bf 12}$ & ${\bf 256}$ & ${\bf 58} \pm {\bf 16}$ & ${\bf 3.8}$ 
\ \ \ \ \ \ \ \ \ \  \\
\hline
${\bf 13}$ & ${\bf 200}$ & ${\bf 36} \pm {\bf 14}$ & ${\bf 2.7}$ 
\ \ \ \ \ \ \ \ \ \  \\
${\bf 14}$ & ${\bf 256}$ & ${\bf 28} \pm {\bf 15}$ & ${\bf 2.0}$ 
\ \ \ \ \ \ \ \ \ \  \\
$15$ & $196$ & $18 \pm 13$ & $1.4$ 
\ \ \ \ \ \ \ \ \ \  \\
$16$ & $201$ & $-18 \pm 11$ & $-1.6$ 
\ \ \ \ \ \ \ \ \ \  \\
\hline
\end{tabular}

\noindent The regions, shown in Fig.~\ref{gridfig}, are numbered from East to
West, looping North to South. 
Cells with a
statistical significance of detection greater than $2 \, \sigma$ are shown in
bold.
Errors are $1\,\sigma$ and on the total (background + net) counts. 
\label{gridtab} 
\end{table}

The analysis of Table~\ref{gridtab} shows that
$7$ regions over the $16$ selected are detected at a $2 \, \sigma$ level
above the background. We count regions from East (left) to West (right) and
North (top) to South (bottom).
Fig.~\ref{gridfig} shows that the regions of excess counts are
concentrated East, West and South of the four bright galaxies of the group: 
regions $9$, $13$ and
$14$ in the East contain $114$ net counts, while
regions $8$,
$11$, and $12$, contain $149$ net counts.
Moreover, half of the cells in
the grid (if we don't take into account cell $6$, detected at a $1.8 \,
\sigma$ level) are compatible with no excess emission over the background,
and most 
of these are North of the four galaxies (regions $1$, $2$, $3$, $4$, $5$,
$10$, $15$ and $16$).  
Hence, the diffuse emission is located in two
distinct regions as well as around the galaxies. 
This could be seen in PBEB's Fig.~2, 
but they gave no quantitative analysis of the spatial distribution of
photons, except for their radial surface brightness profile.

Is it possible
to detect more precisely these excess 
counts regions, without degrading the S/N ratio?
Indeed, the grid we used was arbitrarily set on
the HCG~16 image, and some cells, especially cells $10$, $11$, $13$ and $14$,
overlap 
two regions where the photon densities differ. 
Using a smaller grid would not help because we are at the limit of
sensitivity.
Since we are principally interested in mapping the \emph{diffuse
intergalactic gas}, we can smooth the
image, looking for large-scale features. But we need to be careful with
the level of smoothing: indeed, regions with low count numbers must be
smoothed on larger scales than the bright regions ({\it e.g. \/,} the bright
galaxies),
so as to keep a good
statistical significance of the regions of diffuse emission. Consequently, we
choose an 
adaptive filter algorithm, which automatically adapts the smoothing length
to the local density of photons.
PBEB already used adaptive smoothing to detect diffuse gas in HCG~16 (see
their Fig.~2), but
did not make a quantitative use of the information obtained with this
technique. 

\subsection{Adaptive filtering of the image}

\subsubsection{Spatial resolution}
\label{resol}

In a classical (top-hat) smoothing filter, the filtering radius is fixed, 
the smoothed intensity is the mean (unweighted) counts within this radius,
and the total 
counts within this radius varies across the image.
Hence, every pixel will have a S/N directly proportional to the local
photon density. The pixels in low surface brightness regions will then have
poor statistics. 

In an adaptive filter, the filter size is adapted to contain a fixed number
of counts.
Let
$C_0$ be this fixed number of (background + net) 
counts per smoothing region of size $P$ pixels.
The intensity of a given pixel
will typically be $I = C_0/P$, hence $S/N = I/\delta I = P / \delta P$.
Now, if the cumulative 
counts rise with radius $1\,\sigma$ faster than on average, they
will reach $C_0$ at $P - \delta P$, where the typical counts are $C_1 =
I(P-\delta P)$. 
Since the count curve of growth, $C(P)$, is a Poisson process, one has
$C_0 = C_1 + C_1^{1/2}$. 
Then, in the limit $\delta P \ll P$, one obtains $\delta P =
(P/I)^{1/2}$, hence $S/N = C_0^{1/2}$ (we check that $\delta P/P = C_0^{-1/2}
\ll 1$).
Therefore, \emph{every pixel will
have the same S/N}.  

\begin{figure*}[bt]
\rotatebox{-90}{\resizebox{!}{0.45\hsize}{\includegraphics{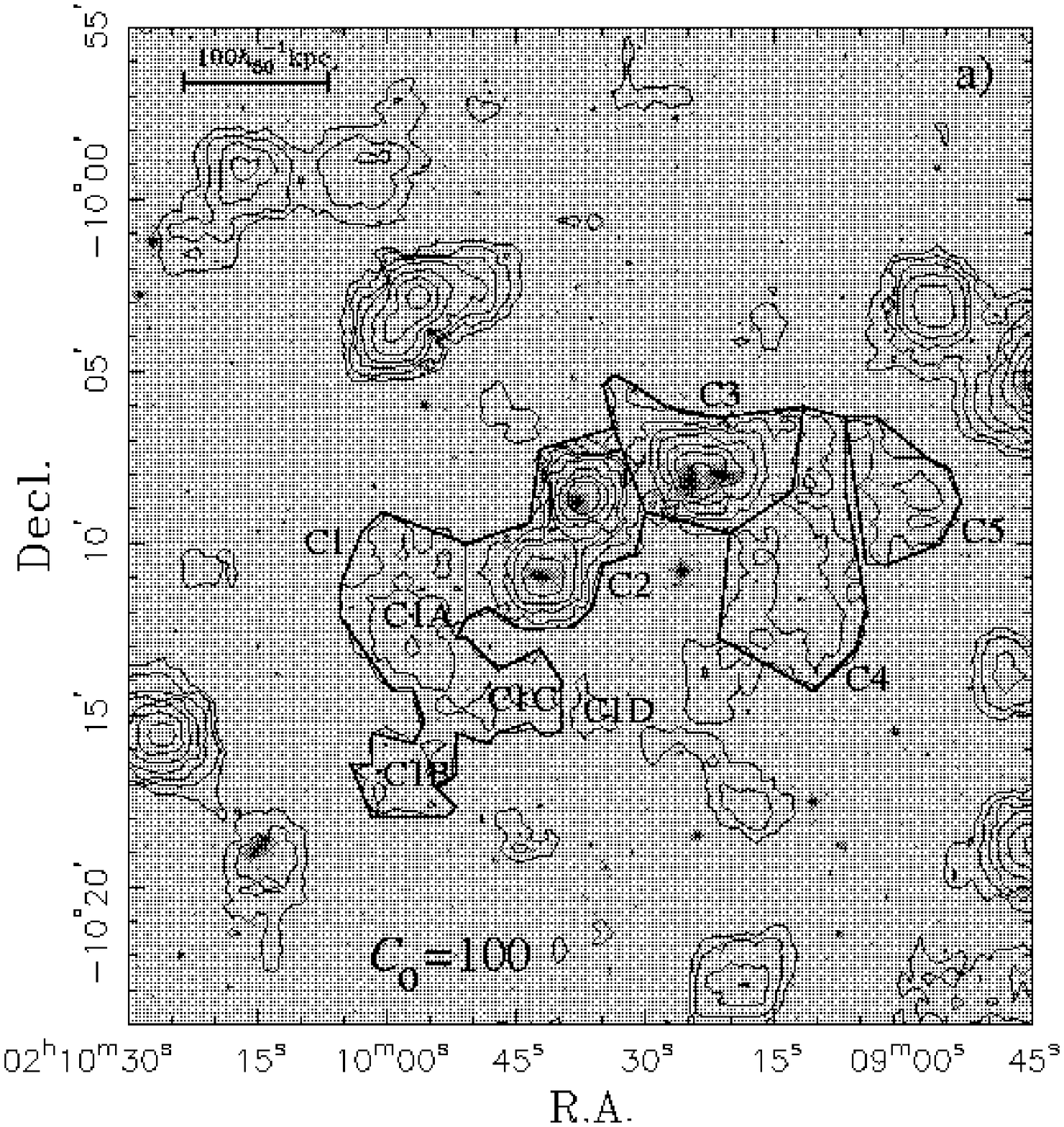}}}
\rotatebox{-90}{\resizebox{!}{0.45\hsize}{\includegraphics{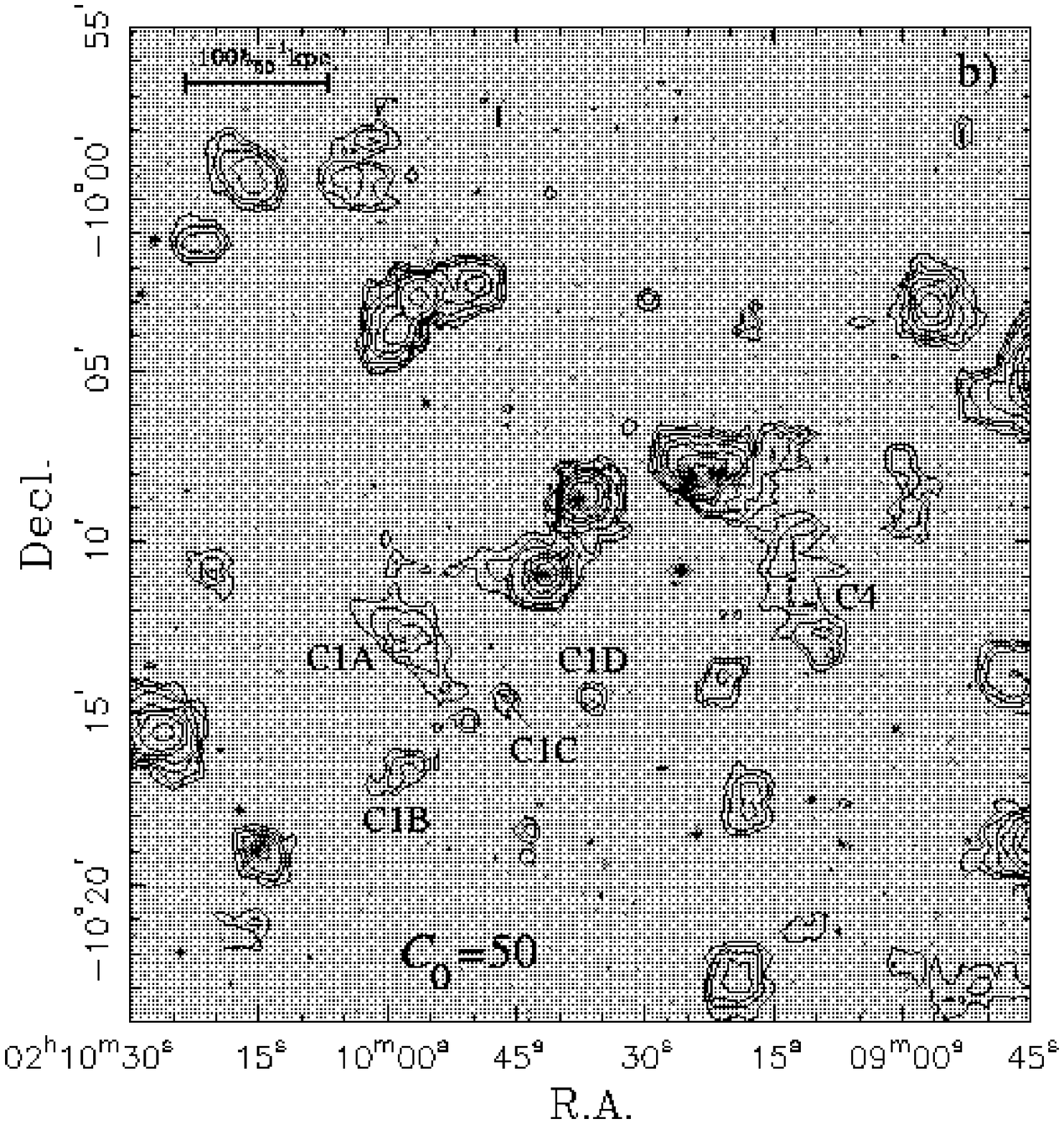}}}
\rotatebox{-90}{\resizebox{!}{0.45\hsize}{\includegraphics{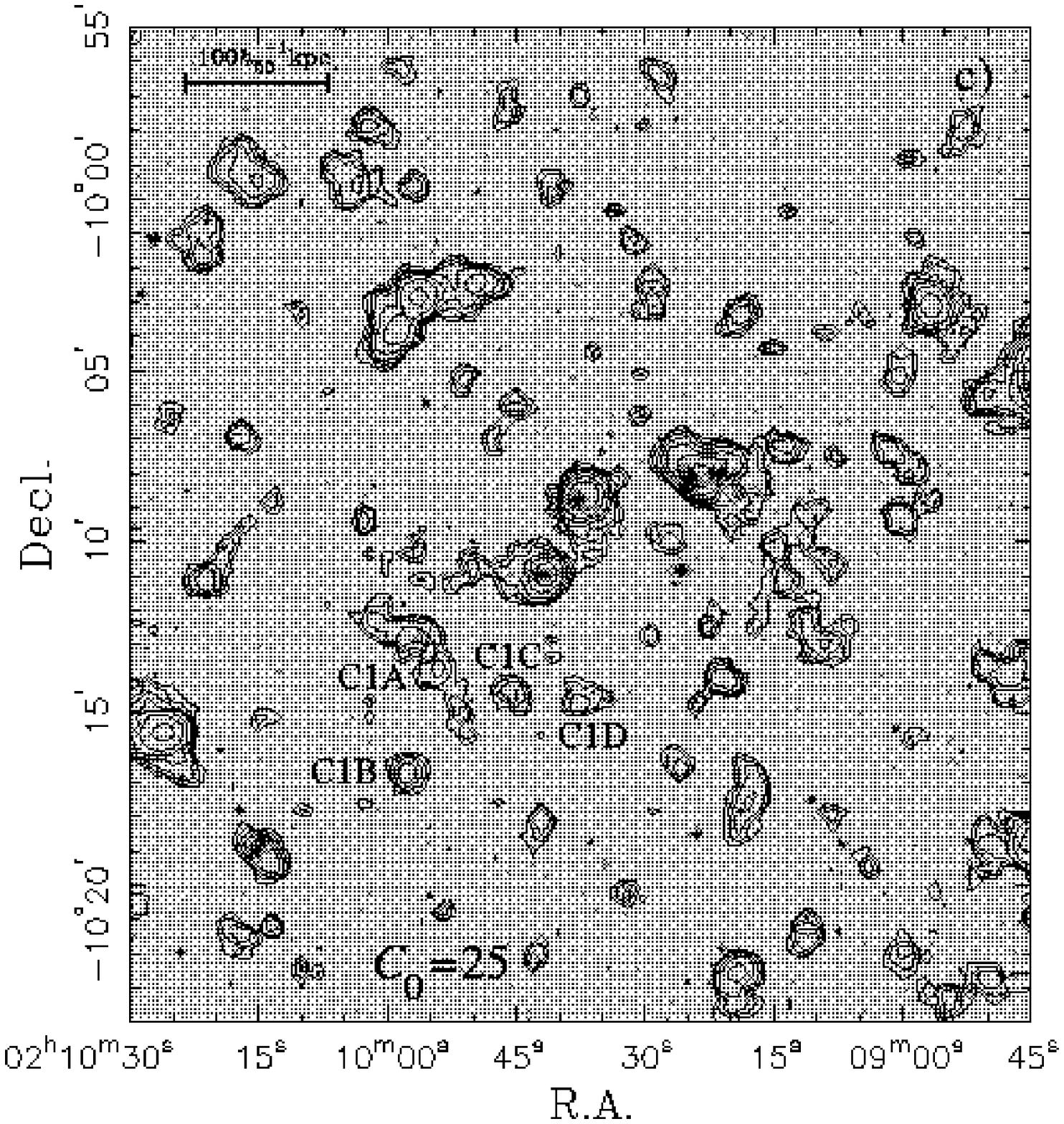}}}
\hfill\rotatebox{-90}{\resizebox{!}{0.45\hsize}{\includegraphics{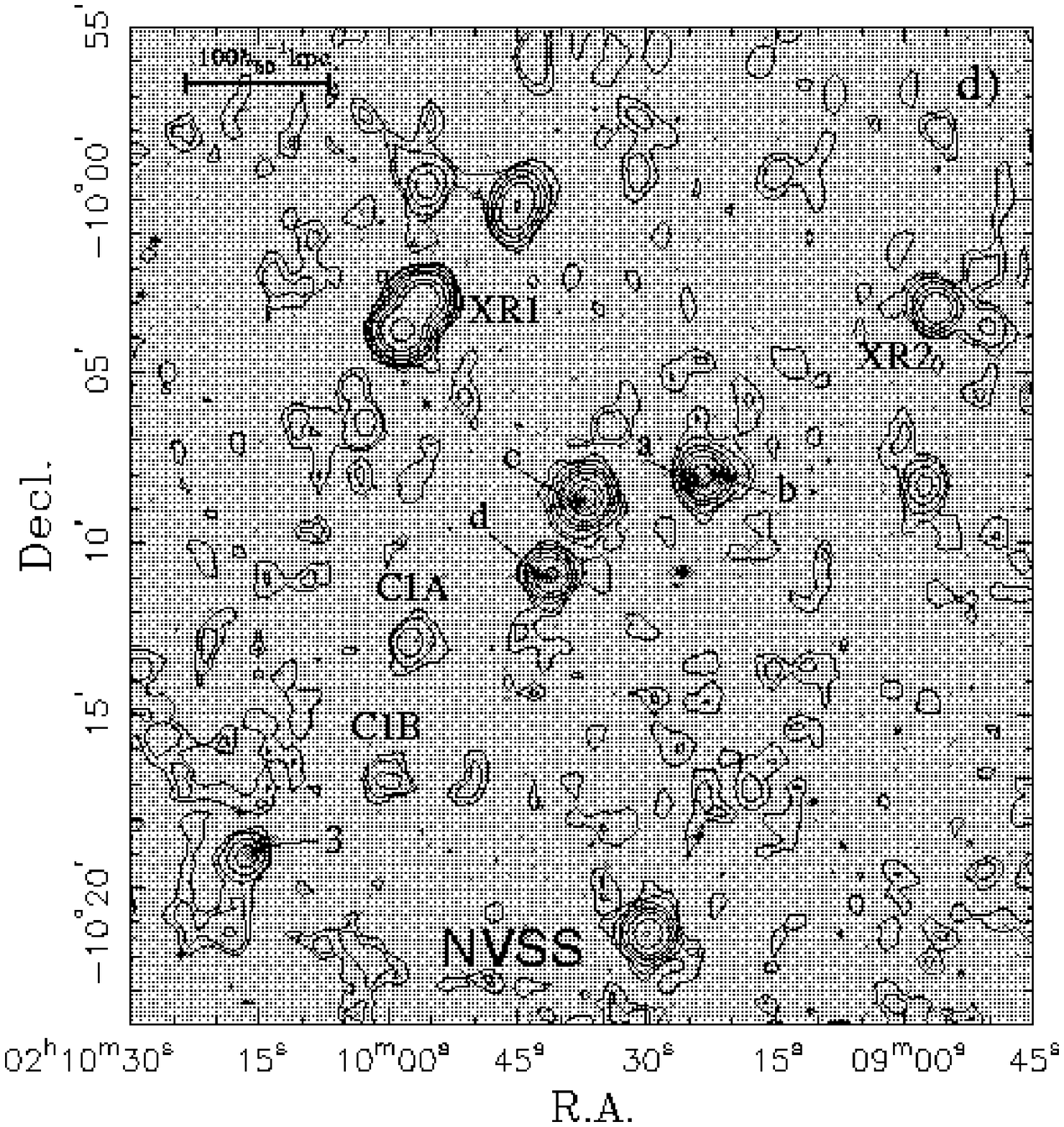}}}
\caption{Contour maps of the adaptively smoothed X-ray emission ($C_0 = 100$
(a), 50 (b), and 25 (c),
see Table~\ref{tabrmax} and Eq.~\ref{eq:rsm})
superimposed on an optical {\sf DSS} image. The five polygonal regions (C1--C5)
dividing 
the emission region in HCG~16 are also shown, as well as the different
components of region C1 (C1A, C1B, C1C and C1D, see text).
Lower right plot (d) shows a
contour map of a 20 cm radio image from the {\sf NVSS} \citep[see
][]{Condon+98}. 
}
\label{adsm}
\end{figure*}

Refining the grid analysis, we define polygonal regions
with much greater S/N than in the grid cells. Contrary to the grid, there
will be no regions half overdense and half-underdense in photons.
We produce adaptively smoothed
images using {\sf ADAPT} in the {\sf PSPC Extended Source
Cookbook} \citep{SMBM94}, with $C_0 = 25$, 50 and 100.
Before smoothing, we masked point sources detected at the $3 \, \sigma$ level,
so as to better highlight the diffuse emission. 
The adaptively smoothed images are shown in Fig.~\ref{adsm}.
For esthetical reasons, the point sources detected by {\sf DETECT} are not
cut in these images.

Our smoothing resolution is easily computed.
Calling $B$ the mean background, since $I > B$, 
our smoothing radius,
equal to $(P/\pi)^{1/2}$, must be smaller than 
\begin{equation}
R_{\rm sm} = \left (C_0\over \pi B
\right)^{1/2}  \ . 
\label{eq:rsm}
\end{equation}
With a mean background of ${B} = 0.565 \, {\rm counts/pixel}$ (see
Sect.~\ref{bgest}), we obtain the smoothing resolutions listed in
Table~\ref{tabrmax}.

\begin{table}[ht]
\caption{Adaptive smoothing parameters}
\begin{tabular}{ccccl}
\hline
$C_0$ & \multicolumn{3}{c}{$R_{\rm sm}$} & \multicolumn{1}{c}{S/N} \\
\cline{2-4}
(cts) & (pixels) & (arcmin) & ($h_{50}^{-1} {\rm kpc}$) & \\
\hline
$#25$ & $3.8$ & $0.9$ & $22$ & #5.5 \\
$#50$ & $5.3$ & $1.3$ & $31$ & #7.6 \\
$100$ & $7.5$ & $1.9$ & $43$ & 10.5 \\
\hline
\end{tabular}

The smoothing radii are obtained through Eq.~(\ref{eq:rsm}) with
a mean background of ${B} = 0.565 \, {\rm counts/pixel}$ (see
Sect.~\ref{bgest}).
\label{tabrmax} 
\end{table}

The grid we used in 
Sect.~\ref{grid} was made of $16 \times 16$ pixel cells. 
Table~\ref{tabrmax} shows that the
lowest resolution varies with $C_0$ from 1/4 square cell ($R_{\rm sm} \sim 4$
pixels)
for $C_0 = 25$ to 1 square cell ($R_{\rm sm} \sim 8$ pixels) for $C_0 =
100$.
This shows that our adopted values of
$C_0$ are well suited for the size of the group, and will not smear out
intermediate scale 
structures by smoothing on too large a scale. 

\subsubsection{Results}
\label{netcounts}



Consider first the image smoothed with $C_0 = 100$ (Fig.~\ref{adsm}a),
which has the largest features of the three images. 
Within the single contour encompassing the group galaxies are
two overdensities, SE and SW 
of the group. They correspond respectively to part or all of cells
$9$, $13$, $14$ for the SE overdensity and 
cells $8$, $11$, $12$ and $15$ for the SW one, in the grid of 
Sect.~\ref{grid}.  
These cell numbers correspond to the cells detected at a level higher than
$2 \, \sigma$ (see Table~\ref{gridtab}). We divide this closed region 
in $5$ different
polygonal regions (C1--C5) filling the contours of the X-ray image adaptively
smoothed with $C_0 = 100$, as shown in Fig.~\ref{adsm}. 
The exact cut between
different regions is arbitrary, but we tried to be consistent with what we
know about HCG~16: C1 is the South-Eastern overdensity, C2 and C3 
are the regions around galaxies c \& d and a \& b of the group (but with
these galaxies cut out), C4 is
the South-Western overdensity, and C5 is a somewhat disjointed region, to the
West of C4. 
The source to the South of C4 is detected as a point-like X-ray source, and
also appears as a double point-like radio-source (Fig.~\ref{adsm}d).
Finally, C6 is the region within the $8'$
radius circle minus the closed region (C1+C2+C3+C4+C5), 
{\it i.e} the part of
this circle ($53 \%$ of it) with no obvious excess counts over the background.


Table~\ref{smoothcounts} shows the net
counts in each region.
Regions C1 and C4 are detected
at $\sim 5 \, \sigma$, much higher than regions C2 and C3 surrounding the
bright galaxies (detected at $\sim 3 \, \sigma$). This shows that the
adaptive smoothing has
targeted regions of high S/N ratio more efficiently than the spatial
detection on the square grid.
But the most
important feature, which fully justifies the use of adaptive filtering, is
that the emission of
region C6 is fully compatible with the background.
Hence,
\emph{half of the region within
$8 '$ of the group center contains negligible emission from 
diffuse hot gas}. 
In Sect.~\ref{shape} below, we confirm the different nature of the two halves
of the  $8'$ circle.

\begin{table}[ht]

\caption{Net counts in regions defined after
adaptive filtering}
\begin{tabular}{lccc}
\hline
Region & Pixels & Net counts & Significance ($\sigma$) \\  
\hline
C1 & $#447$ & $96 \pm 19$ & $5.1#$ \\
C2 & $#261$ & $44 \pm 14$ & $3.2#$  \\
C3 & $#206$ & $45 \pm 13$ & $3.5#$ \\
C4 & $#306$ & $83 \pm 16$ & $5.2#$ \\
C5 & $#175$ & $28 \pm 11$ & $2.5#$ \\
C6 & $1697$ & $10 \pm 31$ & $0.32$ \\
\hline
\end{tabular}

The counts are computed from the unsmoothed images, within the $0.2-2.0\,\rm
keV$ energy interval and 
are rounded to the nearest integer.
Errors are $1\,\sigma$ on the total (background + net) counts.
\label{smoothcounts} 
\end{table}

\subsubsection{Detailed spatial analysis}

\label{detailedadsm}

Nothing has yet been said about the nature of the X-ray photon overdensities
detected. We have intentionally called them with the vague denomination 
``regions''. Are these photons emitted by diffuse gas linked to the group? If
so, is this gas primordial, or mainly ejected by the galaxies? Or
does the diffuse emission originate from foreground or background sources, 
not necessarily linked with the group?
Indeed, looking at PBEB's Fig.~$2$, the position angle of the northern
overdensity seems to be well-correlated with the alignment of three optical
sources.
The spatial and spectral capabilities of  ROSAT are certainly insufficient to
answer these questions, but the careful analysis of the
images and spectra can provide some useful insight. 
Observations of HCG~16 at other wavelengths can also help to specify the
dynamical state of the gas. In particular, deep optical images and radio
surveys can precise the interactions between the diffuse gas, the galaxies and
the radio sources. 

We can now estimate the number of independent smoothing regions within our
regions of excess counts.
If our region has area $A$ pixels and $E$ excess counts,
the number of independent smoothing
regions within it is
\begin{equation}
N = {A\over P} = {A B + E \over C_0} \ .
\label{eq:nindep}
\end{equation}
Using Table~\ref{smoothcounts} and Eq.~(\ref{eq:nindep}), we find that,
for $C_0 = 100$ (our 
worst spatial resolution),
our regions C1, C2, C3, C4, and C5 consist of 3.6, 2.0, 1.7, 2.7,
and 1.3
independent smoothing regions, 
respectively.
Hence, the correlation among neighboring pixels
introduced by adaptive smoothing may connect several local maxima in the
X-ray surface brightness map, which are close in the plane of the sky, 
but not necessarily linked, neither among them nor with the group. 

This is illustrated by comparing the adaptively smoothed images with
different parameters $C_0$. Comparing Figs. \ref{adsm}a and 
\ref{adsm}c,
(with $C_0=100$ and $25$ respectively), 
we see that region C1 is composed of an elongated
structure, almost perpendicular to the group (marked C1A), together with
three point sources 
(marked C1B, C1C and C1D). This substructure remains in
remains at $C_0 = 50$, but is smoothed out in Fig.~\ref{adsm}a 
into the entire C1 
region. All these structures can be seen in Fig.~\ref{gridfig}, although
with a worse S/N.
On the contrary, region C4 seems to be extended, even with the
lightest smoothing. 

\subsubsection{Spatial positioning of the images}


We have obtained an optical image of HCG~16 from the {\sf Digitized Sky
Survey (DSS)}, as well as a $20\,\rm cm$ continuum radio image from the {\sf
NRAO VLA Sky Survey (NVSS)}, which we superpose on a ROSAT {\sf PSPC}
image. The FWHMs are $\sim 2\arcsec$ ({\sf DSS}), $30 \arcsec$ ({\sf PSPC})
and $45 \arcsec$ ({\sf NVSS}). The radio and X-ray images both have $15
\arcsec$ pixels, while the optical image has $1.8 \arcsec$
pixels. Fig.~\ref{adsm}d shows the {\sf NVSS} radio contours overlaid on a
DSS optical greyscale image. Both optical and radio images were centered at
the center of the X-ray image. To ensure a correct superposition, we aligned
the three images using the bright sources detected in the three wavebands.
Comparison of Figs.~\ref{adsm}b,c and \ref{adsm}d clearly shows that 
\emph{the X-ray and 20$\,$cm morphologies are remarkably similar.}

In their spectroscopic survey of HCG~16, \cite{RdC3Z96}
found that six galaxies among the seven belonging to the dynamical
group were emission-line galaxies, AGN, LINERs or starburst galaxies. These
galaxies are good candidates for detection in all three wavebands. Indeed, we
find four sources detected in the three images: they are all HCG~16 galaxies
and are marked a, b, c and d in Fig.~\ref{adsm}d. Galaxies HCG~16a and
HCG~16b seem to share a common X-ray and radio halo, thus
enhancing the probability of tidal interaction between these galaxies
suggested by optical tails seen in galaxy HCG~16a. The differences between
the optical, radio and X-ray positions are less than $5 \arcsec$, much less
than the radio and X-ray PSFs. Moreover, we find two objects exactly
coincident in radio and X-rays, ensuring good correspondence between X-rays
and radio images independently. Both were first detected in the X-rays by
\cite{SC95}, who called them XR1 and XR2, with no optical counterparts.
We found two radio sources
within a radius of less than $5 \arcsec$ around them, and XR1 has \emph{the
same} bimodal structure in X-rays and in radio.

The probability of finding an {\sf NVSS} source within a given radius around
an 
arbitrary position is less than $10^{-3}$ when this radius is $\sim 5
\arcsec$ \citep[see][]{Condon+98}. This means that the coincidences between
six sources in the whole image cannot have occurred by chance, 
and this ensures that the
superposition of the three images is quite perfect, given the positioning
uncertainties. We are now able to
compare small-scale structure in the three wavebands and attempt to elucidate
the 
nature of the three regions of excess X-ray counts.

\subsection{Finalizing the regions of diffuse emission of HCG~16}
\label{diffuseregions}

The problem is now to separate local maxima due to
interloping X-ray sources (point or extended sources) from those due to the
presence of diffuse gas. The regions C2 
and C3, surrounding the four group galaxies, are certainly related to the
group.
We must then analyze the \emph{small scale X-ray structure} of regions C1, C4
and C5 and compare it to optical and radio images.  For this, we use a wavelet
transform of the image to search for structures simultaneously at all scales. 

\subsubsection{Wavelet transform of the X-ray image}
\label{wavesect}
Since its invention in the early 80s, the wavelet transform (hereafter, WT)
has proven its capabilities in numerous astronomical applications, such as
the detection of the large scale structure ({\it e.g.\/,} \citealp{SdLB93}),
galaxy detection and counts \citep{SBM90},  
and structure detection in low-intensity X-ray images \citep{SP98}.
We used the {\sf TRANSWAVE} \emph{\`a trous} \citep{Shensa92,SM94}
wavelet package kindly provided
by E. Slezak.
In the \emph{\`a trous} implementation of the discrete WT,
an $N \times N$ 
image is transformed into $i$ wavelet planes (hereafter WPs) of $N \times N$
pixels, each being the difference between two consecutive 
wavelet smoothings at scales $i$ and $i+1$ (with $2^i$
and $2^{i+1}$ pixels respectively). The pixel values
in these planes are called the wavelet coefficients at scale $i$. The main
advantage of this algorithm is that each WP has the same number
of pixels, and thus, the reconstruction of the image (for example, after
thresholding in the wavelet planes) is a straightforward
process of addition. 
However, there is redundancy in the full set of wavelet coefficients.
The main difficulty with the \emph{\`a trous} wavelet filtering 
is the estimation of the statistical
significance of the pixels in each WP, which should not follow Poisson nor
gaussian statistics.

\begin{figure*}[bt]
\rotatebox{-90}{\resizebox{!}{0.45\hsize}{\includegraphics{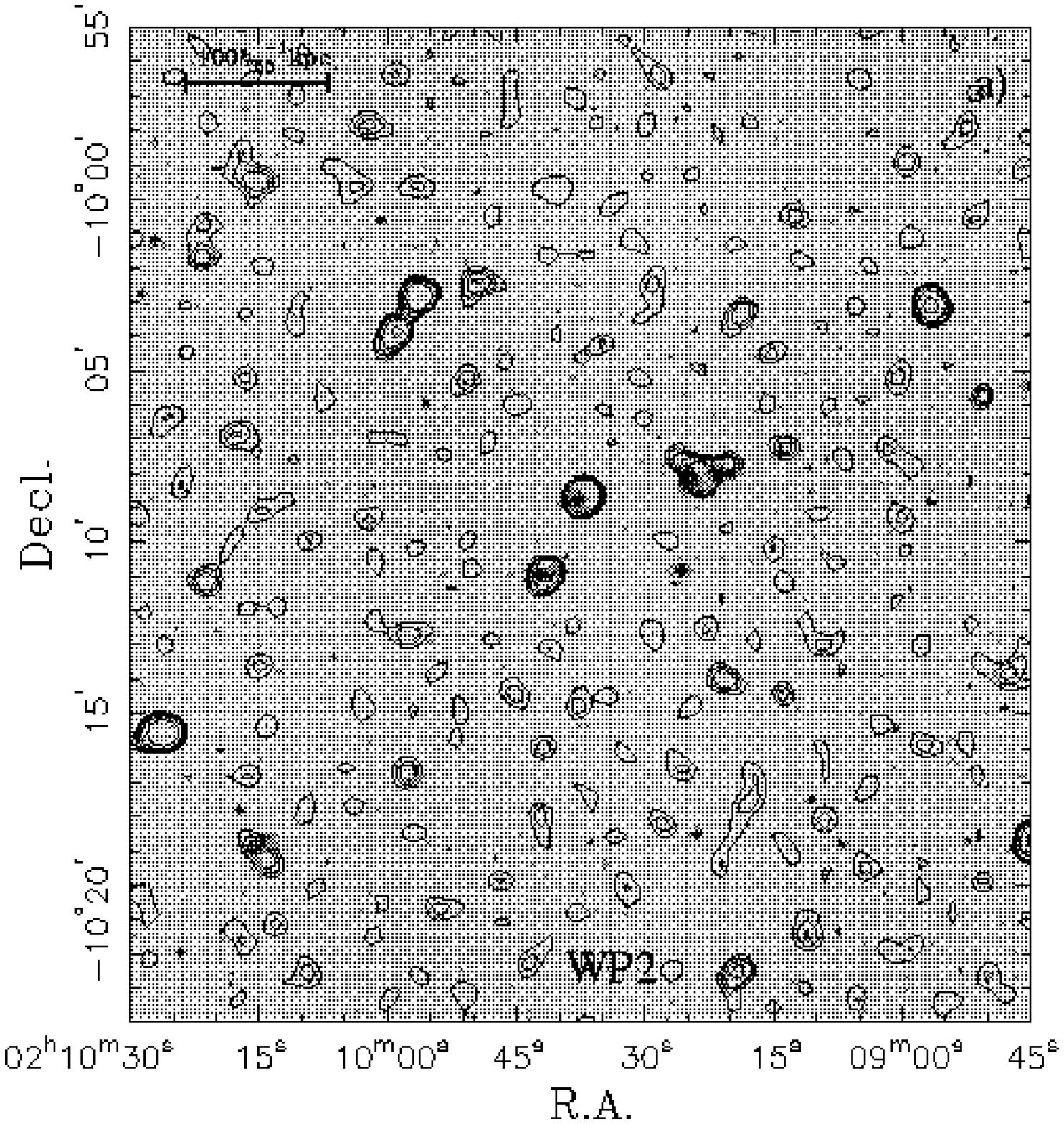}}}
\hfill
\rotatebox{-90}{\resizebox{!}{0.45\hsize}{\includegraphics{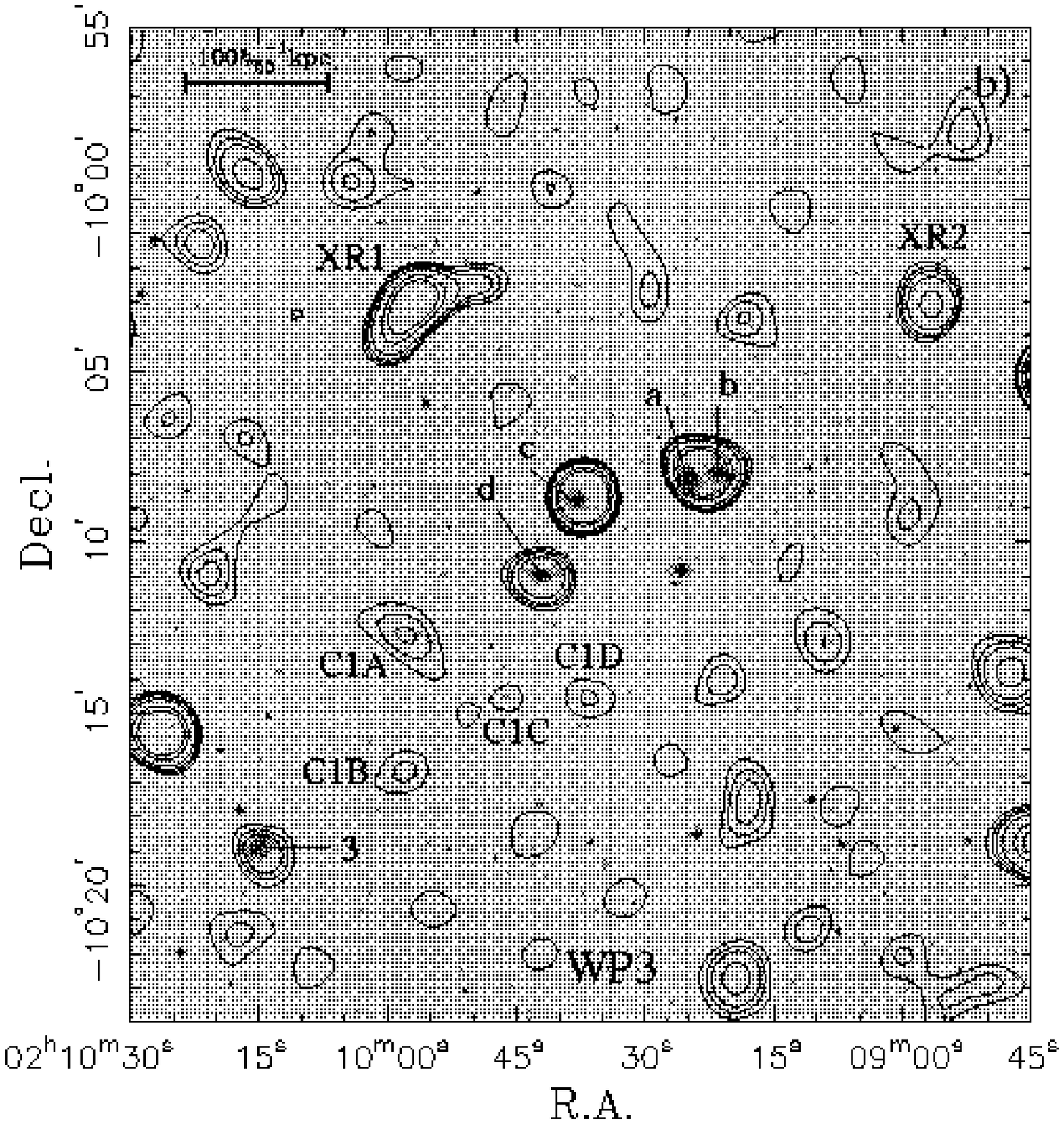}}}
\rotatebox{-90}{\resizebox{!}{0.45\hsize}{\includegraphics{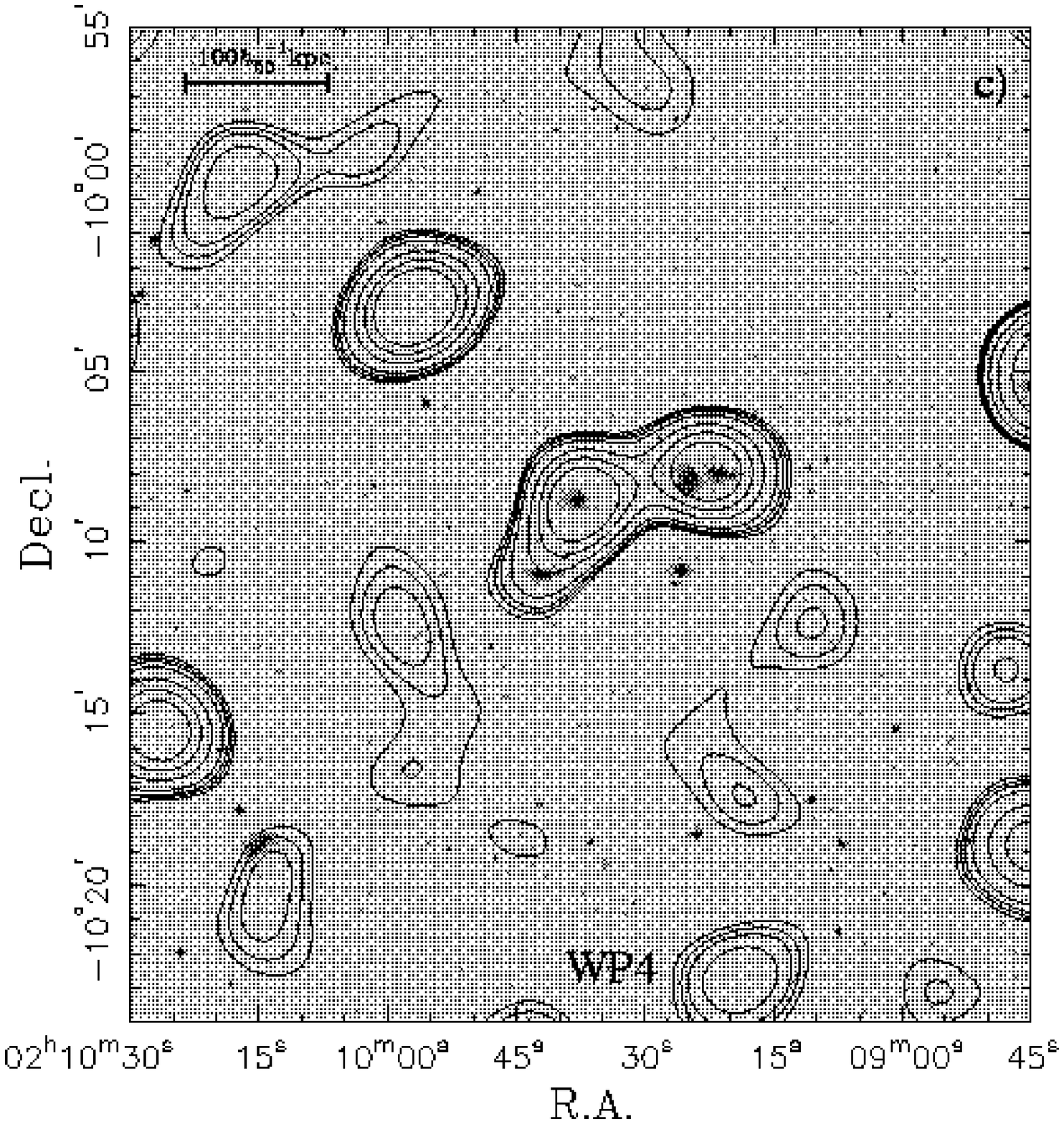}}}
\hfill
\rotatebox{-90}{\resizebox{!}{0.45\hsize}{\includegraphics{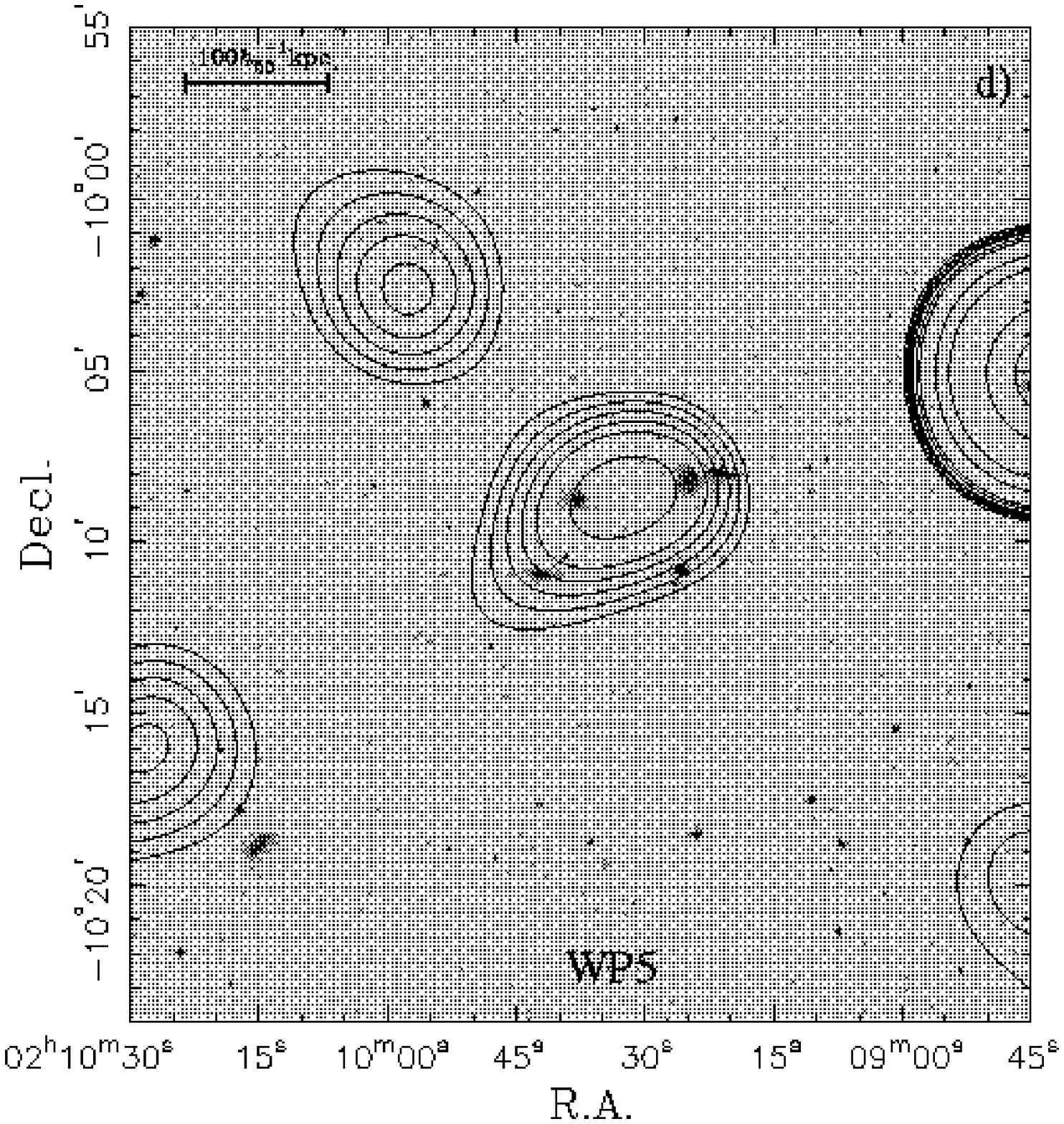}}}
\caption{Contour map of the wavelet planes WP2 (a), WP3 (b), WP4 (c), and WP5
(d)  
superimposed on an optical {\sf DSS}
image. Contours are drawn at $1\, \sigma, 2\, \sigma, 3\, \sigma, 5\, \sigma,
10\, \sigma$ and $50\, \sigma$, where $\sigma^2$ is the $3-\sigma$ clipped
variance of 
the whole WP (see Sect.~\ref{wavesect}). Structures in WP2, WP3 and WP4
are resolved at 1, 2, 4, and $8 \arcmin$ (23, 46, 92, and $184 \, h_{50}^{-1}
\,{\rm kpc}$), respectively.
In Fig.~\ref{wp}b,
the objects XR1 and XR2, detected by
\cite{SC95} are also shown, as well as the different components of region C1.} 
\label{wp}
\end{figure*}





Fig.~\ref{wp}
shows the wavelet
contours of WPs 2, 3, 4 and 5, respectively.
We do not take WP1 into account, since it is highly contaminated
by small-scale noise in the original image, nor WP6 
and higher, which smooth the group on too large a scale to prove useful. 
Structures seen only at the $1\,\sigma$ level are should mostly be artefacts.


There are two ways we can
ensure that a detected source is real: First, if a source is detected in more
than one WP, its probability of being true is enhanced. Indeed, random noise
is not correlated between WPs
and cannot produce large spatial overdensities. Moreover,
it cannot produce pixels high enough to leave some power on larger
scales. Second, the probability of a
false  X-ray detection being randomly
superimposed 
on an optical or radio source is very low.  Consequently, we will search for
small-scale X-ray sources in regions C1, C4 and C5, detected in several WPs and
roughly superimposed on optical and/or radio sources. 

\subsubsection{Nature of region C1}

\begin{figure}[bt]
\rotatebox{-90}{\resizebox{!}{0.9\hsize}{\includegraphics{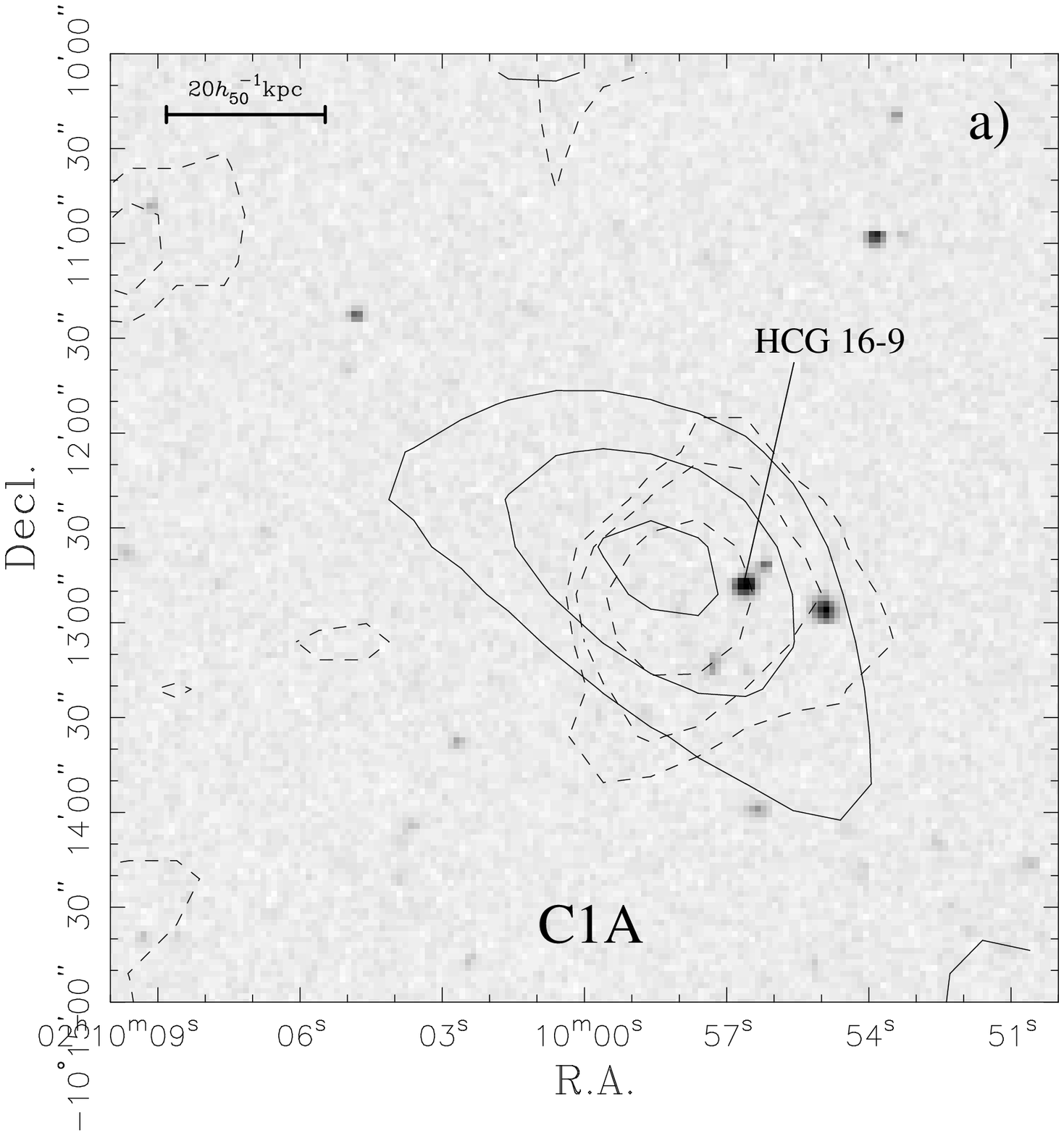}}}
\rotatebox{-90}{\resizebox{!}{0.9\hsize}{\includegraphics{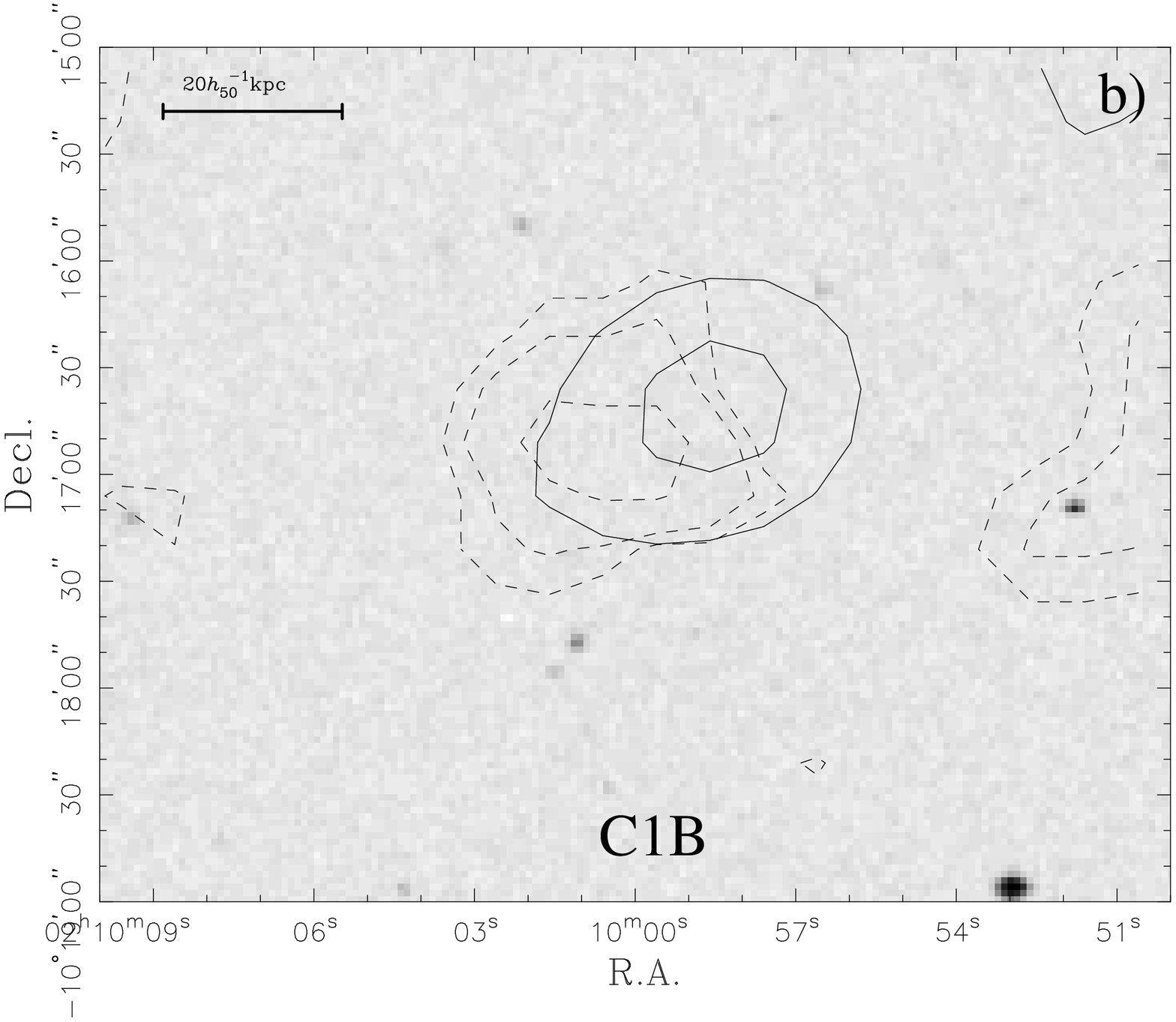}}}
\caption{Zoomed optical images of C1A (a) and C1B (b), together with WP3
X-ray ({\it solid\/}) contours
and {\sf NVSS} radio ({\it dashed\/}) contours. Note the superposition of the
X-ray and 
radio sources and in (a) 
the quasi-perfect alignment of their axis angle with the
direction defined by the two optical sources. 
The brighter one, HCG 16--9, is 
a background galaxy \citep{RdC3Z96}.
The scale (\emph{upper left})
corresponds to the distance of the HCG~16 group.}
\label{c1a}
\end{figure}

The multiple nature of region C1 (Sect.~\ref{detailedadsm}) 
is confirmed by the inspection of the
Wavelet Planes. Four of the 
sources present in WP2  are also detected in WP3 (marked C1A, C1B, C1C and
C1D in Fig.~\ref{wp}b). The first two are still detected in WP4, where the
smoothing is so strong that only an extended region remains,
comparable to C1, with two local maxima (at C1A and C1B) and an evidence of
distortion by C1C.
Note that sources C1A and C1B were detected at
$\sim 3.5 \, \sigma$ and cut using the local detection algorithm {\sf DETECT}
provided in the {\sf PSPC Extended Source
Cookbook}. Thus, the excess photons measured in
Sect.~\ref{netcounts} do not take these sources into account. But, it is
very 
likely that the majority of the diffuse gas is related to these sources.

We now examine sources C1A, C1B, and C1C and compare with 
their optical and radio
counterparts:
\begin{itemize}

\item {\bf Source C1A:} 

Fig.~\ref{c1a}a shows a zoomed optical image of C1A, together
with WP3 contours (\emph{solid}) and radio contours (\emph{dashed}).
There is an {\sf NVSS} radio source less than $10 \arcsec$ away from
this X-ray source. Both sources are coincident with
two optical sources. Moreover, in Fig.~\ref{c1a}a, we verify that the
western part of the radio source is aligned with the elongated X-ray source
and with the direction defined by the two optical sources.
The second of these optical
sources is an
emission line galaxy, HCG~16--9, 
observed spectroscopically by RdC3Z. Its 
recession velocity was measured as more than $20\,000 \,\rm km\,s^{-1}$,
{\it 
i.e. } \emph{it is a background galaxy superimposed on the group}. 
The other optical source is comparable in size and magnitude to HCG~16--9, 
and although we don't
have any spectroscopic evidence for this, it is unlikely that this galaxy
lies at the redshift of HCG~16.

The coincidence of an X-ray, a radio and two optical sources, with one having
a radial velocity 
$16\,000\,\, \rm km \, s^{-1}$ larger than the group makes it very
unlikely that C1A is linked to the group. The radio source denotes a point
source, but the X-ray emission seems to be extended.  It is
difficult to say much more about this source. If it is at the distance of 
HCG~16--9 ($z \sim 0.072$), its extension is several hundreds of
kpc. It may be a background X-ray group with two prominent galaxies.

\item {\bf Source C1B:}


Fig.~\ref{c1a}b shows
that
an {\sf NVSS}  radio source is overlapping the X-ray
extended structure, at $30 \arcsec$ from the
X-ray local maximum. 
The probability that this superposition is random is less
than $0.01$, without taking into account the X-ray position uncertainty
\citep[see][]{Condon+98}. No optical source corresponding  
to C1B is detected either in the DSS or in \cite{dCRZ94}. Here again, the
superposition of a radio and an X-ray source (with no optical counterpart this
time) puts serious doubts on the link between C1B and the group and on the
diffuse nature of C1. 
Moreover, only $10\pm8$ net counts originate from C1B (after the point source
is masked).

\item {\bf Source C1C:}

A total of $33\pm 13$ net counts are emitted from region C1C, which is detected
as a local maximum in WP2 and WP3.  
No radio counterpart is
detected, but a star is found $17''$ from the X-ray peak
(Figs. \ref{adsm}b and 
\ref{wp}b), whose magnitude is $B_J = 16.66$ according to the ROE/NRL {\sf
COSMOS} UKST Southern Sky Object Catalog.
Given that according to the {\sf COSMOS} catalog there are $N=17$ brighter
stars in a $16'\times16'$ box centered on HCG~16, Poisson statistics yield a
probability of $P = 1 - \exp \left [ \pi N (\theta/16')^2\right ] = 1.7\%$ of
having 1 or more stars within $17''$ from the center of C1C.
We thus conclude that the X-ray emission of C1C is linked to a foreground
star.
The $L_X/L_{\rm opt}$ ratio (as defined by \citealp{Motch+98}) is 0.01,
i.e. in the upper range for stars (see Fig.~3 of \citeauthor{Motch+98}).
Note that its X-ray emission was not excised because it was just below the
$3\,\sigma$
excision threshold.

\item {\bf Source C1D:}

$15\pm8$ net counts (8\% of the counts in regions C2-C3-C4, and 16\% of the
counts in C1, see
Table~\ref{smoothcounts})
are detected in region C1D, which has
no radio nor optical counterpart.
We do not incorporate this source in our spectral analysis, but do include it
in our  total group diffuse luminosity.

\end{itemize}

In summary, of the four components of C1, one (C1A) is linked to a
background galaxy and a radio-source, one (C1B) is a radio-source, one
(C1C) is a foreground star, and one (C1D) contributes marginally at best
to the group diffuse emission. Therefore, \emph{at best, only one-sixth of
the X-ray emission from C1 is connected to diffuse gas
in HCG~16.}

\subsubsection{Nature of regions C4 and C5}

The same analysis can be repeated with regions C4 and C5 (regions C2
and C3 are directly around the galaxies of the group, and the excess photons
are likely to be produced in the group). 

\begin{itemize}

\item {\bf Region C4:}
  
Region C4 has an extended structure with three prominent local
maxima, even in Fig.~\ref{gridfig}, where the FWHM of the smoothing is
only $45 
\arcsec$. The three aligned maxima are still apparent at $C_0 = 25$
(Fig.~\ref{adsm}c) and somewhat at $C_0 = 50$.
(Fig.~\ref{adsm}b). 
They are also
recovered in WP2.  
A double structure remains in WP3 and is finally merged in
WP4. No radio or identified optical source is found superimposed
on C4. This reinforces its diffuse nature, without ensuring that this
excess is linked to the group, even if this seems a reasonable
assumption. Indeed, the two interacting galaxies HCG~16a and 16b 
lie at
$\sim 140 \,
h_{50}^{-1} {\rm kpc}$ in projected distance from C4.
This gas could have been ejected by these
galaxies (see Sect.~\ref{disc}).

\item {\bf Region C5:}

\begin{figure}[bt]
\rotatebox{-90}{\resizebox{!}{0.85\hsize}{\includegraphics{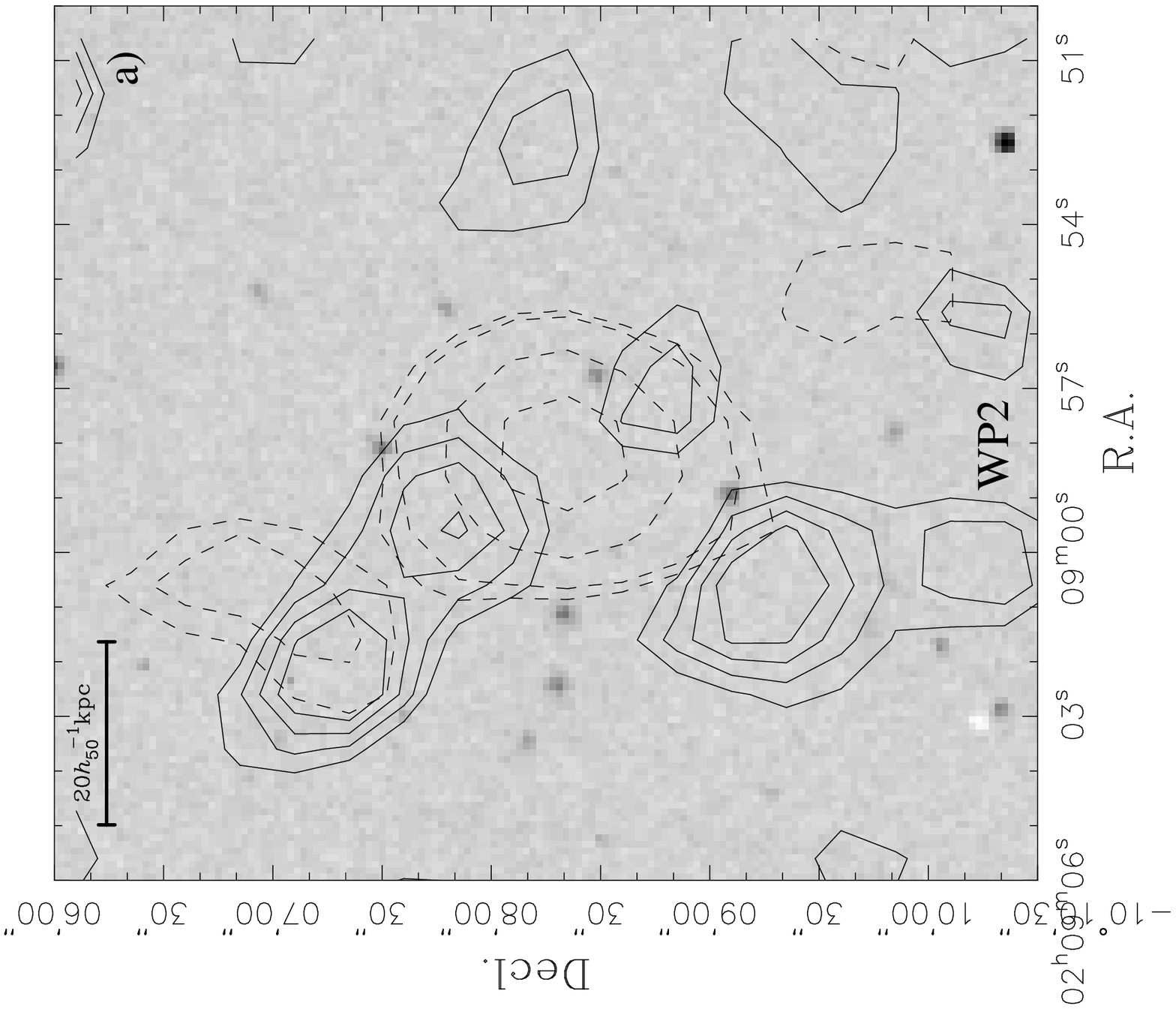}}}
\rotatebox{-90}{\resizebox{!}{0.85\hsize}{\includegraphics{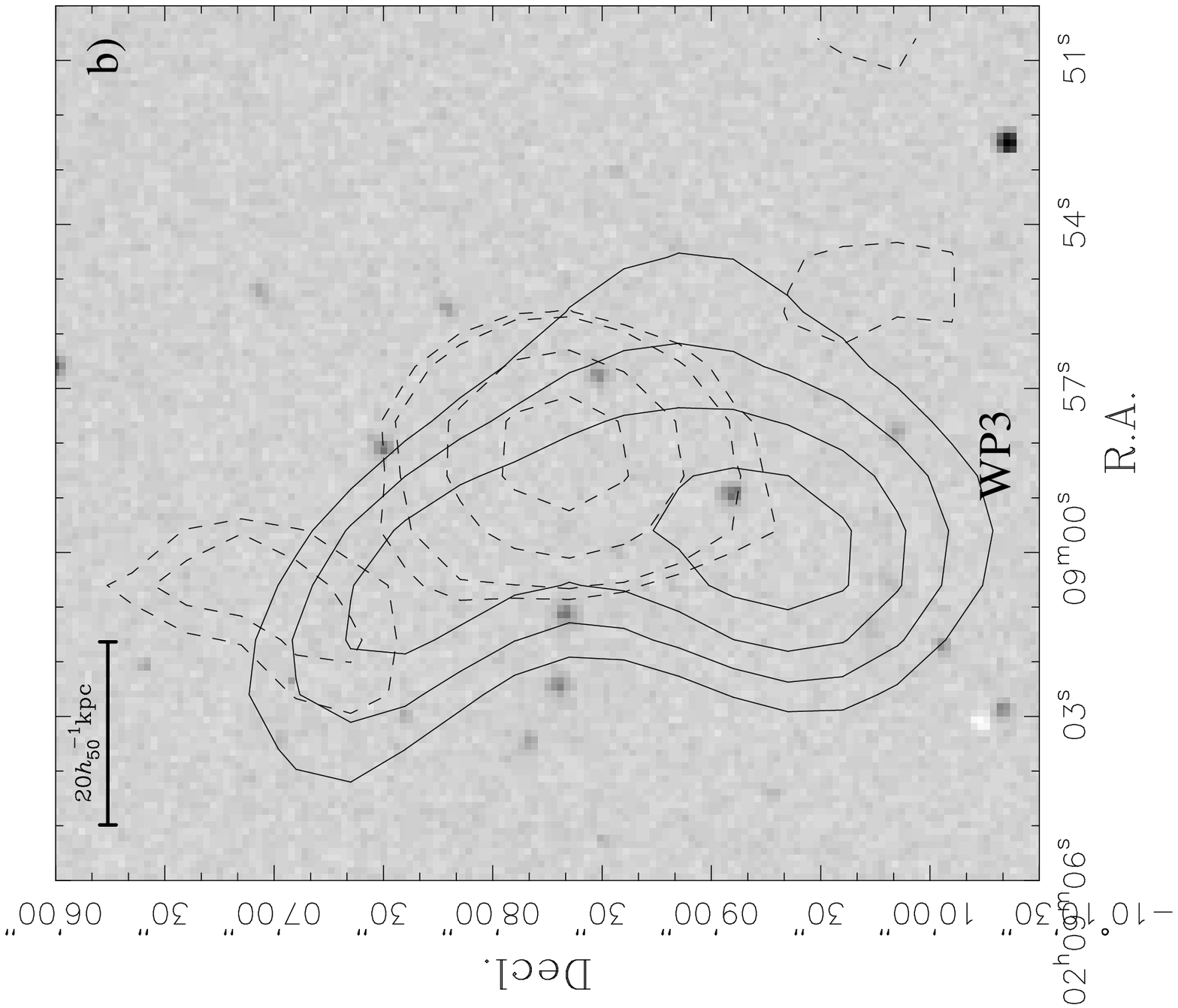}}}
\caption{Radio (\emph{dashed}) and WP2 (a), WP3 (b)
wavelet plane (\emph{solid}) contours of C5 
superimposed on an
optical image. Note the several maxima comprising the X-ray emission, and the
numerous unidentified optical sources present in this area.
}
\label{c5a}
\end{figure}


Figs. \ref{c5a}a and b show the superposition of optical, radio and
 X-ray wavelet planes WP2 and WP3, respectively. Region
C5 has a complex X-ray structure, and
several local maxima can be seen in WP2, which merge into a single
structure in WP3. An $\rm 8.2\,mJy$ {\sf NVSS} radio source is superposed
with the  extended structure in 
WP3, and the prominent three X-ray sources are situated at distances
of 35, 80, and $95''$ from the peak of the radio emission.
Given cumulative source counts of $10^6$ and $4\times10^5$ for 1.4 GHz radio
fluxes greater than 5 and $15\,\rm mJy$ respectively \citep{Condon+98},
and the 82\% {\sf NVSS} coverage of the celestial sphere, we infer 20.7
{\sf NVSS} sources brighter than $8.2\,\rm mJy$ per square degree.
Therefore, Poisson statistics yield respective probabilities of 0.6\%, 3.2\%
and 4.5\%  of having each of 
the three X-ray sources  as close as they are to the
radio-source.

Likewise, at least six
optical sources are clearly visible within the outer WP3 isophotes of C5, of
which two are {\sf COSMOS} galaxies brighter than $B_J = 20.1$, and there is
an additional {\sf COSMOS} galaxy lying just $10''$ outside of the outer WP3
isophote (to the East).
Within the $16'\times16'$ box centered on the group, there are 34 galaxies in
COSMOS with $B_J \leq 20.1$.
Poisson statistics then yield a 3.6\% probability of having as least 3 {\sf
COSMOS} galaxies within $10''$ of the WP3 outer contour (in a $5.5\,\rm
deg^2$ region).

If these 3 galaxies were at the distance of HCG~16, their typical separations
would be $20 \, h_{50}^{-1} \, \rm kpc$ and they would all be less luminous
than the SMC. We would then have a very compact subgroup of very faint dwarfs
detached from Hickson's original compact group of 4 bright galaxies.

The alternative of a background group or cluster appears much more plausible,
given that subgroups of very faint dwarfs have never been discussed, and that
there is X-ray emission apparently associated with this group or cluster.

Therefore, the
concordance of the radio source with 3 X-ray sources, \emph{coupled with}
the large galaxy surface number density in C5, strongly suggests 
that C5 is not associated with diffuse emission from HCG~16.

\end{itemize}

\subsection{Summary}

We have shown that the detection of diffuse gas within a radius of $\sim 200
\, h_{50}^{-1} {\rm kpc}$ around HCG~16 reduces to the significant detection of
five regions of diffuse X-ray emission, 
filling less than half of the
circle. The small-scale structure of these regions, together with
associations with optical and radio point sources, allows us to reject two of
these regions (C1 
and C5) as
related to point sources or background extended sources. There thus remains
three 
diffuse emission regions, two (C2 and C3) surrounding the four bright galaxies 
and the other (C4)
$\sim 140 \, h_{50}^{-1} {\rm kpc}$ away from the two 
interacting
galaxies HCG~16a and HCG~16b. 
A better understanding of the gas physics in these regions requires an
examination of their spectra.

\section{Spectral analysis}
\label{spectral}

The spectra are extracted using the
{\sf QPSPEC} task in the {\sf IRAF/\-PROS} environment. 
Again, we
restrict our analysis to the energy range $0.20 - 2.01 \,\rm
keV$ unless otherwise noted.
We first compare the shapes of the spectra (before background subtraction)
of different regions with that of
the background.

\subsection{Spectral shapes}
\label{shape}


The shapes of the spectra, $C_k$, in specific regions are compared to
the shape of the background spectrum, $B_k$, renormalized to the total counts
of the spectrum of the region of interest. 
For this, we use likelihoods with Poisson statistics:
\begin{eqnarray*}
\ln {\cal L} =  \sum \! C_k \left [\!\ln \!\left (\!{\sum C_k \over \sum B_k}
\!\right ) 
\!-\! 1 \right ] 
+ \sum \! C_k \ln B_k - \!\sum_k \sum_{j=2}^{C_k} \!\ln j \ ,
\end{eqnarray*}
where $k$ is the spectral channel.
%
We resort to sets of $10\,000$ 
Monte-Carlo trials
to derive probabilities that the $\ln {\cal L}$s of simulated background
spectra drawn from the observed background, measured relative to the
normalized background spectrum
are smaller than
the observed $\ln {\cal L}_{\rm obs}$ 
measured between the
spectra of a given region and of the normalized background.
The simulated spectra are drawn from a Poisson distribution to match the
normalized 
background spectrum in each energy channel, i.e. with expected simulated
counts  $B_k \left (\sum C_k / \sum B_k \right)$.

\begin{table}[ht]

\caption{Comparison of spectral shapes  with the background spectral shape}
\begin{tabular}{lccl}
\hline

Region & Total counts & $\ln {\cal L}$ & \multicolumn{1}{c}{$P (\ln {\cal L} < \ln {\cal L}_{\rm obs})$} \\
\hline
C2+C3+C4 & 663 & $-77.7$ & \ \ \ \ \ \ \ \ 0.026 \\
C6 & 974 & $-77.1$ & \ \ \ \ \ \ \ \ 0.258 \\
\hline
\end{tabular}

The fourth column 
represents the probability (from Monte-Carlo simulations) that the region has
a spectral shape 
consistent with that of the background.
\label{stattest} 
\end{table}

Table~\ref{stattest} shows the resultant probabilities that each spectral
region has a spectrum with the same shape as that of the background.
Not only does the region of excess emission, C2+C3+C4 present an excess of
counts, but it also has a   
spectrum whose shape is significantly different from that of the background, 
while region C6 with
negligible \emph{net} counts has a spectral shape consistent with that of the
background.
Hence, \emph{the spectral shape analysis confirms that the
regions of excess counts are indeed the locations of X-ray emission and not
caused by poor background subtraction.} 


\subsection{Spectral fits}

The background-subtracted, point-source excised spectra were then fit to a
hydrogen absorbed 
\citep{BCM92} {\sf MEKAL} 
(\citealp{MGO85,MLO86}, 
with Fe L calculations by \citealp{LOG95} and the ionization balance from
\citealp{AR85}) plasma,
using {\sf XSPEC} version 10, with $\chi^2$ minimization.
%
A 
vignetting correction of the background was performed before the
back\-ground-subtracted spectrum was analyzed. 

Table~\ref{galaxies} shows the parameters we used for extracting the galaxy
spectra.
Table~\ref{specfits} presents the results of our spectral fitting to
different regions within HCG~16.
There remains many instances where energy channels have fewer
than 10 net counts, and the Poisson statistics do not resemble gaussians,
hence our $\chi^2$ spectral fits are not fully appropriate.
Therefore, the reduced $\chi^2$ values and the
90\% error bars given in Table~\ref{specfits} should not be over-interpreted.

\begin{table}[ht]
\caption{Extraction parameters for galaxy spectra}
\begin{tabular}{lccc}
\hline
Galaxy	&	RA	&	Dec	& radius \\
\cline{2-3}
	& \multicolumn{2}{c}{(J2000)}	& (arcmin) \\
\hline
HCG~16a\&b & 
$2^{\rm h} 09^{\rm m} 24 \fs 0$ & $-10 \degr 07 \arcmin 49''$ & 1.5 \\
HCG~16c & $2^{\rm h} 09^{\rm m} 38 \fs 2$ & $-10 \degr 08 \arcmin 49''$ & 1.2
\\
HCG~16d & $2^{\rm h} 09^{\rm m} 43 \fs 9$ & $-10 \degr 10 \arcmin 58''$ & 1.1
\\
HCG~16--3 & $2^{\rm h} 10^{\rm m} 16\fs 5$ & $-10 \degr 19 \arcmin 11''$ &
1.1--0.9 \\
\hline
\end{tabular}

The spectrum of HCG~16--3 was extracted from an ellipse.
\label{galaxies}
\end{table}

\begin{table*}[ht]
\caption{Spectral fits}
\tabcolsep 3pt
\begin{tabular}{lcccccccccc}
\hline
Region & Energy range & Net & $kT$ & $Z$ & $N_H$ & $h_{50}^2\,\rm VEM$
& $\chi^2$ &  
$h_{50}^2\,L_X^{[0.5-2.3]}$ & $h_{50}^2\,L_{\rm bol}^{\rm unc}$  & 
$h_{50}^2\,L_{\rm bol}$ \\
\cline{9-11}
 & (keV) & Counts & (keV) & ($Z_\odot$) & ($10^{20} \rm cm^{-2}$) 
& ($10^{63}\rm cm^{-3}$) & (/d.o.f.) &
\multicolumn{3}{c}{($10^{41}\,\rm erg\,s^{-1}$) }
\\
\hline
C2+C3+C4 & 0.20--1.40 & 125 & $0.27^{+0.28}_{-0.10}$ & (0.05) &
galactic &  31.4 & 0.64 & 0.39 & 0.68 & 2.13 \\
	&	&	& $0.24^{+0.17}_{-0.07}$ & 0.1 & galactic & 24.5 &
0.61 & 0.43 & 0.69 & 2.32 \\
	&	&	& $0.19^{+0.11}_{-0.06}$ & 1 & galactic & #4.5 & 0.69
&  0.48 & 0.71 & 3.31  \\
	&	&	& $0.19^{+0.10}_{-0.06}$ & 10 & galactic & #0.5 &
0.71 & 0.48 & 0.72 & 3.69 \\
\\
C4 & 0.20--1.31 & #71 & $0.26^{+31.7}_{-0.11}$ & 0.1 & galactic & 11.9 & 0.30
& 0.23 & 0.36 & 1.03 \\
	& & & $0.20^{+0.27}_{-0.07}$ & 1 & galactic & #2.3 & 0.33 & 0.26 &
0.38 & 1.53 \\
	& & & $0.20^{+0.22}_{-0.07}$ & 10 & galactic & #0.3 & 0.33 & 0.27 &
0.38 & 1.69 \\
\\
HCG~16a\&b & 0.20--2.01 & 232 & $0.72^{+0.24}_{-0.19}$ &
$0.17^{+0.43}_{-0.09}$ & galactic & 16.5 & 0.58 & 0.97 & 1.21 & 1.99 \\
	&	&	& $0.72^{+0.25}_{-0.18}$ & 0.1 & galactic & 21.8
& 0.63 & 0.96 & 1.27 & 2.30 \\
	&	&	& $0.70^{+0.25}_{-0.23}$ & 1 & galactic & #4.0 &
0.74 & 0.90 & 0.99 & 1.40 \\
\\
HCG~16c & 0.20--2.01 & 213 & $0.53^{+0.16}_{-0.15}$ &
$0.79_{-0.55}^{+\infty}$ & galactic & #7.3 &  0.39 & 0.93 & 1.01 & 1.42 \\
	& 	&	& $0.65^{+0.19}_{-0.13}$ & 0.1 & galactic & 20.3
& 0.90 & 0.86 & 1.13 & 2.00 \\
	&	&	& $0.53^{+0.17}_{-0.14}$ & 1 & galactic & #4.0
& 0.37 & 0.94 & 1.02 & 1.42 \\
\\
HCG~16d & 0.20--2.01 & #83 & $0.67^{+\infty}_{-0.31}$ & 0.1 & galactic &
#7.2 & 0.32 & 0.31 & 0.41 & 0.71 \\
	&	&	& $0.39^{+\infty}_{-0.14}$ & 1 & galactic & #1.5 &
0.27 & 0.31 & 0.35 & 0.53 \\
	&	&	& (0.12) & 0.1 & (41) & (123) & 0.19 & 0.14 & 0.16 &
190 \\
	&	&	& (0.11) & 1 & (44) & (240) & 0.19 & 0.14 & 0.16 &
1770 \\
\\
HCG~16--3 & 0.20--2.01 & #32 & (0.49) & 0.1 & galactic & #2.3 & 0.23 & 0.08 &
0.11 & 0.21 \\
	&	&	& (0.29) & 1 & galactic & #0.6 & 0.21  & 0.10 & 0.12
& 0.23 \\
\hline
\end{tabular}

\noindent 
Column (1): region.
Column (2): energy range used for the spectral fit.
Column (3): net counts in the given energy band (hence lower than the
net counts in Table~\ref{smoothcounts}).
Column (4): temperature.
Column (5): metal abundance.
Column (6): absorbing hydrogen column density (the galactic value is $N_H =
2.02 \times 10^{20}\,\rm 
cm^{-2}$, \citealp{Stark92}).
Column (7): volume emission measure, defined as
$\int_{V} n_{e}  n_{p}  dV$.
Column (8): reduced $\chi^2$ (per degrees of freedom) of fit. Their values
are low because the noise is not gaussian.
Column (9): X-ray luminosity in the $0.5 - 2.3\,\rm keV$ band.
Column (10): bolometric X-ray luminosity, uncorrected for absorption.
Column (11): bolometric X-ray luminosity, corrected for absorption.
The error bars are 90\% confidence levels for one interesting parameter.
Values in parentheses indicate provide the best fit when the fit was
unconstrained ({\it i.e.\/,} when the 90\% 
confidence levels could not be determined).
Values of the metal abundance without error bars nor parentheses were frozen
in the fit. 

\label{specfits}
\end{table*}

The low net counts in our background-subtracted spectra
(Table~\ref{specfits}) make it difficult to
perform reliable spectral fits for temperature, metal abundance, and
absorbing column. Indeed, decent spectral fits require at least 500 net
counts, whereas we have 
typically 2 to 7 times less.
Nevertheless, the spectra often do have sufficient counts to provide
decent constraints on the gas
temperature, as well as on the bolometric luminosity, once metal abundance
and absorbing Hydrogen column are fixed to 
reasonable values.
Note the large absorption corrections on the bolometric luminosity, especially
for the lower best fit temperatures, where the emission is peaked at low
energies, which are the most seriously affected by absorption.



The temperature of the diffuse emission is well constrained: {\it
the diffuse emission is cool} at $kT = 0.27^{+0.28}_{-0.10}$ (with much
narrower 90\% confidence intervals for fixed metal abundance).
All fits
produce an 
upper-limit (90\% confidence level) of 0.56 keV.
The metallicity is completely unconstrained.

\begin{figure}[bt]
\rotatebox{-90}{\resizebox{!}{\hsize}{\includegraphics{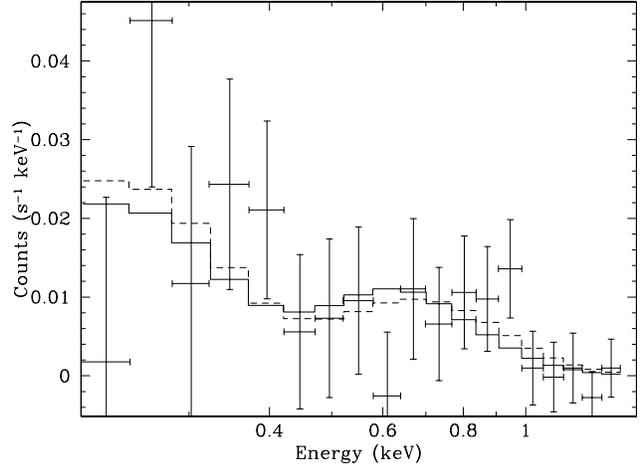}}}
\caption{Spectral fits to the diffuse emission in HCG~16 (regions
C2+C3+C4). The \emph{solid} 
and \emph{dashed histograms}
represent the best fit {\sf MEKAL} models for solar and one-tenth solar
metallicity plasmas, with a galactic absorbing column ($N_H = 2.0 \times
10^{20} \,\rm cm^{-2}$).}
\label{gasell_spec}
\end{figure}

The spectral fits from region C4 (offset from the group galaxies) are similar
to that of the total diffuse emission, and C4 accounts for roughly half
of the total luminosity of the diffuse emission. 
However, C4 could be hot if its metal abundance is low.
Fig.~\ref{gasell_spec} 
shows the best-fit {\sf MEKAL}
spectrum for region C2+C3+C4 together with the residual error per bin.
The fit is adequate, hence no additional galactic absorption nor additional
component is required by the spectrum.

The galaxy pair HCG~16a\&b and galaxy HCG~16c both
emit more X-ray luminosity,
before correction for absorption, than the group, illustrating the difficulty
of absorption corrections in low temperature spectra.

With only 83 photons the spectrum of HCG~16d produces
virtually no constraints on
temperature or metallicity.
Nevertheless, careful inspection of the spectrum reveals that the $6$
first energy channels (between $0.2$ and $0.5 \, {\rm keV}$) have very low net
counts, even taking into account the great error bars, suggesting strong
hydrogen absorption, as confirmed by the best fits with variable absorption,
which produce a column density 20 times the galactic value.
However, the absorption correction to luminosity then becomes
enormous and produces unreasonably high absorption-corrected bolometric
luminosities.

\subsection{Comparison with other X-ray results on HCG~16}

\subsubsection{Group diffuse emission}

The EINSTEIN satellite pointed for $3217 {\rm s}$ at HCG~16 with the {\sf IPC}
detector. 
The analysis of these observations by
\cite{BHR84} led to $26 \pm
12$ counts for the entire group (galaxies plus intergalactic medium), 
corresponding to a luminosity of $2 \times 10^{41} \, h_{50}^{-2} \,{\rm
erg\,s^{-1}}$. 
But the limited sensitivity and angular resolution of the IPC
didn't allow separating the galactic and intergalactic components.

\citeauthor{SC95}'s (\citeyear{SC95}) analysis 
of the ROSAT {\sf PSPC} observation of HCG~16
led to the conclusion that the 
X-ray emission was due to point sources associated to the galaxies and not to a
diffuse intra-group medium. By assuming $kT = 1.0 \, {\rm keV}$ and
$Z = 0.1 \, Z_{\odot}$, they obtained a $5 \, \sigma$ upper
limit on diffuse emission of $3.0 \times 10^{40} \, h_{50}^{-2}\, {\rm
erg\,s^{-1}}$ in the 
$0.5-2.3 \, {\rm keV}$ band, after converting this flux to luminosity correctly (see
footnote in Sect.~\ref{introd} and Table~\ref{specfits}), i.e., 
$\simeq 30\%$ below our fitted luminosity in the
same energy band.\footnote{adopting $kT = 0.3\,\rm keV$ instead of $1.0\,\rm
keV$ 
produces roughly the same upper limit.}

On the
other hand, PBEB, who analyzed the same data as \citeauthor{SC95}, 
but in the full
$0.1-2.4 \, {\rm keV}$ {\sf PSPC} energy band, found an
excess of photons in an $8 \arcmin$ radius circle corresponding to a bolometric
luminosity of $4.8 \times 10^{41} \, h_{50}^{-2}\, \rm erg\,s^{-1}$.
Their factor of 16 discrepancy with \citeauthor{SC95} can be subdivided into
three terms:
1) a factor 3.4 caused by the correction for Hydrogen (and Helium)
absorption,
2) a factor 1.6 caused by the narrower energy band used by \citeauthor{SC95},
and
3) a factor 3.0 unexplained residual term.
The adaptive smoothing of the image
(PBEB's Fig.~2) indeed
showed two diffuse extensions on both sides of the group
galaxies, roughly corresponding  to the regions C1 and C4 defined in
Sect.~\ref{resol}.

Our spatial analysis is performed over an intermediate energy band
($[0.2,2.0] \,\rm keV$, see Sect.~\ref{datared}). 
From our spectral fit to the diffuse emission (regions C2+C3+C4),
we find 
$L_X [0.5-2.3\,\rm kev] = 3.9 \times 10^{40} \,h_{50}^{-2}\,\rm erg\,s^{-1}$,
only slightly 
higher than the upper limit of \citeauthor{SC95}.
Our bolometric luminosity is
$L_{\rm bol} = 2.3 \times 10^{41} \,
h_{50}^{-2}\,\rm erg\,s^{-1}$ (after adding 9\% extra luminosity to include
region C1D, assuming that it has the same spectrum as the rest of the group
emission).
Thus
\emph{our derived bolometric luminosity for the diffuse emission is  
half that of \cite{PBEB96}}.

Summing up from our Table~\ref{smoothcounts}, our $8'$ circle around HCG~16
has 306 net counts.
In comparison, we infer from the surface brightness profile of PBEB (their
Fig.~1) that they measure 281 net counts within a circle of radius 
$0\fdg 14 = 8\farcm 4$ around HCG~16.
In both cases, the group galaxies and point sources exceeding a $3 \sigma$
threshold were excised. Thus, the difference (although within the error bars)
between PBEB's net counts and ours can be attributed to a different
background region. Indeed, their background region was chosen to be an
annulus of radii between $36'$ and $42'$, while ours was taken between
$26'$ and $34'$ (see Table~\ref{BGtab}). We showed in Sect.~\ref{bgest}
that the mean background value increases with off-center radius, which
explains our higher net counts. But this can certainly not explain the
discrepancy in luminosity. Hence, 
\emph{the discrepancy between the high luminosity measured by \cite{PBEB96} and
our low 
luminosity for the diffuse emission is not caused by different background
subtraction or different 
removal of the emission from bright galaxies, but by \citeauthor{PBEB96}'s
inclusion of regions associated with radio or optical galaxies, some of which
are definitely background objects}.

We do agree with PBEB that this diffuse emission is cool, as they find
$kT = 0.3 \, \pm 0.05 (1\,\sigma)\, {\rm keV}$, consistent with our values. 
%
Note that the correction for absorption is as large as 3.5 at temperatures as
low as $0.3\,\rm keV$ and increases rapidly for decreasing temperature.
Hence, while our bolometric luminosity, uncorrected for absorption, depends
little 
on the fit, the absorption corrected luminosity is less certain.

{}From Table~\ref{smoothcounts}, region C6 has $10\pm31$ net counts.
Hence the 90\% confidence ($1.28\,\sigma$) and $3\,\sigma$ confidence upper
limits for diffuse emission in C6 are 50 and 103 net counts, respectively.
We convert these net counts to luminosities assuming the same spectrum as
region C2+C3+C4, which has 172 net counts according to Table
\ref{smoothcounts}.  
After correction
for galactic absorption, this yields $L_{\rm bol}^{\rm C6} < 6.1 \times 10^{40}
h_{50}^{-2} \,\rm erg\,s^{-1}$ (90\%) and $1.3
\times 10^{41} h_{50}^{-2} \,\rm erg\,s^{-1}$ $(3\,\sigma$).
Hence, region C6 contributes little to the diffuse X-ray emission of
the group (unless it is much cooler, hence more absorbed).

\subsubsection{X-ray emission from the group galaxies}

Our $0.5-2.3 \, {\rm keV}$ luminosities for the galaxies match fairly
well those of SC95, who had simply assumed the galaxies to have $1\,\rm keV$
temperature and solar metallicity: our luminosities are 1\%, 12\% and 28\%
lower for HCG~16a\&b, HCG~16c, and HCG~16d respectively.

The extent of the diffuse X-ray
emission  around each galaxy or pair in HCG~16
is fairly large ($ \simeq 40 \, h_{50}^{-1}
\, \rm kpc$) in comparison with the extent of the
X-ray emission detected 
by \cite{HC99} around spiral pairs or by \cite{RP98} around most interacting
pairs. 
Fig.~15 of Read \& Ponman shows that only the pairs closest to maximum
interaction show such large extents of their diffuse 
X-ray emission, but their diffuse-X-ray to optical luminosity ratio are
factors of 5 larger than for the HCG~16ab pair.

\section{Discussion}
\label{disc}


We estimate below 
the total dynamical mass of the group, the
baryonic fraction together with the separation of the baryonic mass into its
different components (namely, hot gas, HI, ${\rm H_2}$ and stars) and the
possibility of virialization of HCG~16.

\subsection{Mass budget of HCG 16}


\subsubsection{Total dynamical mass}
\label{dynmass}

\begin{table*}[hbt]
\caption{Dynamical mass estimates}
\begin{tabular}{lccccccccc}
\hline
Group & $N$ & $\theta$ & $R$ & $\sigma_v$ & 
$M_{\rm vir}^{\rm HTB}$ & $M_{\rm proj}^{\rm HTB}$ & 
$M_{\rm avg}^{\rm HTB}$ & $M_{\rm med}^{\rm HTB}$
& $M_{\rm proj}^{\rm BT}$ \\
\cline{6-10}
 & & (arcmin) & ($h_{50}^{-1} \,\rm kpc$) & ($\rm km \, s^{-1}$) &
\multicolumn{5}{c}{($h_{50}^{-1} \,10^{12} M_\odot$)}\\
\hline
\cite{H82} & 4 & #3.2 & #74 & 99 & 1.1 & 1.1 & 0.9 & 0.7 & 0.4 \\
\cite{RdC3Z96} & 7 & 16.9 & 389 & 76 & 2.0 & 2.0 & 2.0 & 1.9 & 0.8 \\
\hline
\label{massopt}
\end{tabular}

Column (2): number of galaxies.
Columns (3) and (4): radius of the smallest circumscribed circle
containing the $N$ galaxies, in arcmin and in kpc respectively.
Column (5): unbiased sample velocity dispersion (corrected for the
measurement errors, see Eq.~[\ref{sigmav}]).
Columns (6), (7), (8), and (9):
virial, projected, average and median mass
estimates from \cite{HTB85}, respectively.
Column (10):
projected mass estimate for isotropic orbits from \cite{BT81}.
These mass estimates do not include correction for measurement errors, and
are thus slight overestimates.
\end{table*}

The total mass of the group can be estimated, to first order, assuming that
the group is in dynamical equilibrium (we will return below to the relevance of
this assumption).
Table~\ref{massopt} presents the estimates for the total mass of the group
using either the 4 original galaxies \citep{H82} of HCG~16, or adding to
them the
three 
additional galaxies that were found by \cite{dCRZ94} in the close
environment of
HCG~16, and confirmed 
spectroscopically by RdC3Z.
We apply the mass estimates of \cite{HTB85} relevant to
self-gravitating systems as well as the projected mass of \cite{BT81},
relevant  
to test objects orbiting within an underlying potential.
We use the radial velocity $v_i$
measurements and errors $\delta v_i$ from RdC3Z.
For the velocity dispersion, we computed
\begin{equation}
\sigma_v = \left [{1\over N-1} \sum_i (v_i - \bar v)^2 - {1\over N}
\sum(\delta v_i)^2 \right]^{1/2} \ .
\label{sigmav}
\end{equation}

We adopt the median of the four Heisler et al. mass estimates, as we
deem it unlikely that the galaxies are test particles in a potential, since
we do not detect this potential in diffuse X-rays.
Interpolating the total mass at $8'$,
within which we have constraints on the gas content of the group,
then yields $M_{\rm tot} \simeq
1.4\times 10^{12} h_{50}^{-1} M_\odot$.
For a spherical group, this yields a density that is 256 times the critical
density of the Universe or 770 times the mean density if $\Omega_0 = 0.3$.

Of course, the reliable estimation of the total gravitating mass of the group
is difficult with 
only four to seven galaxies.
Moreover, the group may not be in virial equilibrium (see
Sect.~\ref{virgroup}
 below).
\cite{M93_Aussois_Dyn,M95_Chalonge} has quantified the effects of departures
from virial 
equilibrium on the estimation of the masses of groups, initially following the
Hubble expansion, taking into account
the softened nature of galaxy potentials.
If 
the galaxies were point masses, the mass of a galaxy system near full
collapse
should be half of the mass inferred from dynamical equilibrium (hereafter
virial mass), as is well known.
But since galaxies have softened potentials, the velocity at closest approach
is only a little larger than for the future virialized system (before it
coalesces).
So, \emph{
the virial mass should provide an adequate estimate
of the mass --- within the apparent radius of a galaxy system near full
collapse}. 
Hence the masses given in Table~\ref{massopt} are probably roughly correct,
unless projection effects are important in HCG~16.


\subsubsection{Mass in hot gas}

An upper limit to the {\it observed} mass of the diffuse intergalactic
gaseous medium (IGM)  
can be estimated from the normalization of the
{\sf MEKAL} plasma model, which is the
Volume Emission Measure (VEM) defined by
\begin{equation}
{\rm VEM} = \int_{V}n_{e}  n_{p} \, dV
= \overline{n_e n_p} V
\label{vem}
\end{equation}
where $V$ is the volume of the emitting region, 
$n_e$ and $n_p$ are the electron and proton densities.
Including the contribution from Helium, the gas mass is
\begin{equation}
M_{\rm hot} = m_p \overline n_p [1+Y/(1-Y)] V 
\label{mgas}
\end{equation}
where $m_p$ is the proton mass and $Y$ the Helium mass fraction.
Since ${\overline n_p}^2 \leq \overline{n_p^2} = \overline {n_e n_p} /
[1+Y/(2-2Y)]$ (the equality being reached for a plasma of uniform density),  
Eqs. (\ref{vem}) and (\ref{mgas}) lead to
\begin{eqnarray}
M_{\rm hot} &\leq& {m_p V^{1/2} {\rm VEM}^{1/2} \over (1-Y)^{1/2}
(1-Y/2)^{1/2}} \nonumber \\
&=& 1.2 \times 10^{9} \,h_{50}^{-5/2} \nonumber \\ &\times&
\!\! \left [\left ({\theta \over 1'}\right)^3 \!\!-\!\! \left ({\theta_g \over
1'}\right)^3 \right]^{1/2} 
\!\!\left (h_{50}^2 {\rm VEM \over 10^{63}\,{\rm cm^{-3}}}
\right)^{1/2} \!\!\! M_\odot \ ,
\label{mgasmax}
\end{eqnarray}
where we took a distance $D = 79.2\, h_{50}^{-1} \, \rm Mpc$ and $Y = 0.24$,
%
and
where $\theta$ is the angular radius of the extended emission, and $\theta_g$
is the angular radius at which the emission from a possible group galaxy was
cut.

Eq.~(\ref{mgasmax}) is valid for spherical diffuse regions.
Whereas, C3 and C4 are nearly circular, C2 is close to being comprised of
two circles. 
The VEM of C4 is taken from Table~\ref{specfits}, while that of C2 and C3 are
each taken as half the difference of the VEM of C2+C3+C4 in 
Table~\ref{specfits} minus the VEM of C4.
Moreover, both circular regions of C2 are assumed to have the same VEM.
Table~\ref{mgastb} presents the estimates of the upper limit to the
mass in diffuse intergalactic gas
$M_{\rm hot}^{\rm max}$ for the various
regions of diffuse emission.
Summing up the contributions of the different regions of diffuse X-ray
emission, we obtain $M_{\rm hot} < 2.1 \times 10^{10} h_{50}^{-5/2}
M_\odot$.

\begin{table}[ht]
\caption{Upper limits to the observed gas content of diffuse regions}
\tabcolsep 4pt
\begin{tabular}{lccccc}
\hline
Region & $h_{50}^2\,\rm VEM$ & $\theta$ & $\theta_g$ & 
$h_{50}^{5/2} \,M_{\rm hot}^{\rm max}$ & $\rho_{\rm hot}^{\rm max}$\\
\cline{3-4}
	& ($10^{63}\,\rm cm^{-3}$) & \multicolumn{2}{c}{(arcmin)} &
 ($10^{10} \,M_\odot$) & ($h^{-3/2} \rho_c$) \\ 
\hline
C2A  & #4.9 & 1.7 & 1.2 & #0.24 & 212\\
C2B  & #4.9 & 1.7 & 1.2 & #0.24 & 212 \\
C3 & #9.8 & 1.9 & 1.2 & #0.49 & 269 \\
C4 & 11.9 & 2.0 & 0.0 & #1.2# & 412 \\
C2+C3+C4 & 31.5 & & & 2.2 & 309 \\
C6 (face value) & #2.5 & 8.0 & 0.0 & #4.3# & #24 \\
C6 (90\%) & 12.6 & 8.0 & 0.0 & #9.6# & #53 \\
C6 ($3\,\sigma$) & 25.9 & 8.0 & 0.0 & 13.8# & #76 \\

\hline
\end{tabular}

\label{mgastb}

\noindent The last column is the upper limit to the density of hot gas in
units of the critical density of the Universe.
\end{table}

One does not gain much in attempting to fit a $\beta$ model to each diffuse
region. Indeed, if the gas density profile is $n(r) = n_0 /
[1+(\theta/\theta_c)^2]^{1/2}$, as in the isothermal $\beta$ model with $\beta
= 1/3$ \citep[close to the slope found by ][for HCG62]{PB93}, then
Eqs. (\ref{vem}) and (\ref{mgas})
lead to 
\begin{eqnarray*}
{M_{\rm hot} \over M_{\rm hot}^{\rm uniform}}  = {3^{1/2} \over 2} 
{x (x^2+1)^{1/2} - \sinh^{-1} x \over x^{3/2} (x - \tan^{-1} x)^{1/2}} 
\ ,
\end{eqnarray*} 
where $x = \theta / \theta_c$.
$M_{\rm hot} / M_{\rm hot}^{\rm uniform}$
is always greater than 0.87 for
$\theta_c > \theta / 100$.
If, on the other hand, one assumes $\beta = 1$ as found by \cite{MZ98}, 
Eqs. (\ref{vem}) and (\ref{mgas}) 
lead to
\begin{eqnarray*}
{M_{\rm hot} \over M_{\rm hot}^{\rm uniform}}  = 24^{1/2}
{x^{-3/2} \left [\sinh^{-1} x - x/(x^2+1)^{3/2} \right ] 
\over \left [\tan^{-1} x + x/(x^2\!\!+\!\!1) -
2x/(x^2\!\!+\!\!1)^2\right ]^{1/2}} \ ,
\end{eqnarray*}
which falls to 0.47 for $\theta_c = \theta/5$ (close to what we infer
was derived by  \citeauthor{MZ98}, but to 0.25 for $\theta_c = \theta/10$.
Hence, the uniform approximation for the mass of diffuse gas
is probably valid to within a factor of two
or at worst four.
 
An upper limit to the gas mass from region C6 can be estimated,
assuming that it is a sphere of $200 \, h_{50}^{-1} \, \rm kpc$ radius, and
that its temperature and metal abundance are the same as the best fit case
for regions C2+C3+C4.
The upper limits for C6 are provided in Table~\ref{mgastb}, with the
face value counts, the 90\% upper limit or the $3\,\sigma$ upper limit.
The gas mass of C6 is poorly constrained in contrast with its diffuse
X-ray luminosity.
However, Table~\ref{mgastb} shows that
\emph{if the undetected region (C6) has similar temperature, metallicity and
clumpiness as the detected regions (C2+C3+C4), then its 
mass-density is at most
($3\,\sigma$ limit) 
one-quarter that of the detected
regions}.
Similarly, its mean X-ray surface brightness is at most one-sixth that of
C2+C3+C4. 
Hence, \emph{the low emission of C6 is not merely 
a statistical fluctuation in X-ray counts, but indicative of a true
underdensity in the distribution of hot diffuse gas}.

In summary, we obtain four upper limits to the mass in diffuse hot gas within
$8'$ from the group center:
$M_{\rm hot} < 
2.1 \times 10^{10} h_{50}^{-5/2}$ (if the 10 counts in C6 are just noise),
$6.4 \times 10^{10} h_{50}^{-5/2}$ (taking the 10 counts in C6 at face value),
$1.2\times 10^{11} h_{50}^{-5/2}$ (with 90\% upper limit on C6),
and
$1.6 \times 10^{11} h_{50}^{-5/2}$ (with $3\,\sigma$ limit on C6).

\subsubsection{Baryonic fraction}

Within an $8'$ radius from the group center,
the molecular gas mass is 
$M_{\rm H_2} = 7.4\times 10^{10} h_{50}^{-2} M_\odot$ \citep{LCM98}, while
the mass in cold HI gas is slightly less than 
$M_{\rm HI} = 4.5\times 10^{10} h_{50}^{-2} M_\odot$ \citep[][whose
VLA map shows that a small fraction of the diffuse HI emission extends beyond
the $4'$ radius circle around the group]{Williams98}.
Note that the contributions of dust and ionized hydrogen 
(estimated by MPABB), though important in comparison
with other galaxies, are negligible within the mass budget of HCG~16.
Thus, if the 10 counts measured in region C6 are just noise or taken at face
value, the mass in diffuse hot gas is much smaller than the mass in cold gas
and fairly negligible within the total mass budget of HCG~16.



The fraction $f_b$ 
of baryons within $8'$ of the optical center of HCG~16 is
\begin{eqnarray}
f_b &=& 
{M_* + M_{\rm HI} + M_{{\rm H}_2} + M_{\rm hot} \over M_{\rm tot}} \nonumber \\
&=& \left ({M_*/L_B \over 6.4} + 0.085 \right ) h_{50}^{-1} + f_{\rm hot}\,
h_{50}^{-3/2} \ , \\
\label{fbaryonic}
\end{eqnarray}
where $M_*/L_B$ is the mean stellar mass-to-light ratio for the group and
$f_{\rm hot}$ is the mass fraction of the group in hot gas, and is
1.5\%, 5\%, 8\% or 11\% depending on the interpretation of the 10 net
counts in region C6 (noise, face-value, 90\% limit and $3\,\sigma$ limit,
respectively).


Because the galaxies in HCG~16 have probably all undergone fairly recent
bursts of star formation \citep[e.g.][]{RdC3Z96}, their stellar mass to
blue luminosity ratios are probably much lower than for normal spirals ({\it
i.e.} $M_*/L_B < 2.5$).
There are two ways to estimate $M_*/L_B$ for each of the four bright
galaxies in HCG~16.

First, MPABB mentioned that galaxies HCG~16a and 16c have rotation
velocities consistent with the Tully-Fisher (1977, hereafter TF) relation, 
whereas galaxy
HCG~16d has a rotational velocity at most half of what is expected by the
TF relation. 
\nocite{TF77}
This translates to a luminosity that is at least 16 times larger than expected
from the TF relation.
Assuming that such was also the case for galaxy HCG~16b (its low
rotation velocity on one side is also half of what is expected from the
TF relation, although its high rotation
velocity on the other side is consistent with the TF relation), this leads to
$M_*/L_B = 1.19$ assuming $(M_*/L_B)_{\rm normal} = 2.5$ for the
group of 4 galaxies. 


Alternatively, $M_*/L_B$ can be constrained by the
times since the last bursts of star
formation in each of the 4 bright galaxies of HCG~16.
These times can be inferred from the optical colors of the galaxies or from
their X-ray properties.
Table~\ref{colors} provides the values of $M_*/L_B$ inferred from the colors
of the 4 bright galaxies, using Fioc and Rocca-Volmerange's (1997) {\sf
PEGASE} spectral evolution model, assuming a single burst, a \cite{RB92}
 initial mass function, and solar metallicity.

\begin{table}[ht]
\caption{Colors, ages and stellar masses of the HCG~16
galaxies}
\tabcolsep 2pt
\begin{tabular}{lccccccc}
\hline
Galaxy & $B_T$ & $B_T^0$ & $B\!-\!R$ & $(B\!-\!R)^0$ & Age & $M_*/L_B$ &
$h_{50} M_*$ \\ 
	&	&	&	&	& (Gyr) &	& 
($10^{10} M_\odot$) \\
\hline
HCG~16a & 12.99 & 12.76 & 0.82 & 0.72 & 0.50 & 0.39 & 3.0 \\
HCG~16b & 13.74 & 13.27 & 1.07 & 0.86 & 0.68 & 0.52 & 2.5 \\
HCG~16c & 13.40 & 13.10 & 1.04 & 0.91 & 0.71 & 0.57 & 3.2 \\
HCG~16d & 13.90 & 13.42 & 1.29 & 1.08 & 0.95 & 0.75 & 3.1 \\
\hline
\end{tabular}

Column (1): Galaxy name.
Columns (2), (3), and (4): Asymptotic blue magnitude before and after
correction for Galactic and internal extinction, and mean color within the
$\mu_B = 24.5 \rm mag \, arcsec^{-2}$ isophote, all from \citet*{HKA89}.
Column (5): Color corrected for galactic and internal reddening using
$(B-R)^0 = B-R + 0.54 (B_T^0-B_T)$.
Columns (6) and (7): age and $M_*/L_B$
from the {\sf PEGASE}
spectro-photometric evolution model of \cite{FRV97}.
Column (8): Mass in stars.
\label{colors}
\end{table}
The stellar masses in Table~\ref{colors} yield $M_*/L_B = 0.53$ for the group
of 4 galaxies.
With this value of $M_*/L_B$, Eq.~(\ref{fbaryonic}) yields a baryonic
fraction of 18\% or 21\% if the emission from C6 is respectively 
noise or taken at face
value, and as much as 25\% (90\% limit on C6) or 28\% ($3\,\sigma$ limit on
C6).

Note that if significant star formation  occurred before the last burst, the
mean colors of the galaxies would be redder than with the most recent
starburst. 
Therefore,
blue colors indicate even more recent starbursts than listed in
Table~\ref{colors}. 
This in turn leads to lower $M_*/L_B$ and an even lower baryonic fraction.
Moreover, if $M_*/L_B \leq 0.55$, then there is more mass in cold (HI+H$_2$)
gas than in stars.

\subsection{The dynamical state of HCG~16}
\label{dynstate}

The knowledge of the dynamical state of the HCGs is primordial to assess
the reality of these close associations of galaxies on the sky. 
In paper~II, we show that the low velocity dispersion of HCG~16 is indicative
of a non-virialized dynamical state, but show that the alternative scenario
of chance alignments is even less likely for HCG~16.
Here, we just want to briefly emphasize two points:
is the peculiar X-ray morphology found in HCG~16 compatible with
virialization, and are the X-ray properties deduced from {\sf ROSAT\/} data
reduced here compatible with chance alignment models? 
  
\subsubsection{Can a virialized group have an irregular X-ray morphology?}
\label{virgroup}

If HCG~16 were in a (nearly) virialized state,
the galaxy halos should have merged, and thus the global group potential
should be fairly smooth.
Moreover, the diffuse gas associated with
these halos should have also merged, and reached hydrostatic equilibrium
within this smooth potential.
For example, HCG~62 is the archetype of such a virialized group, as its X-ray
morphology is smooth and the diffuse X-ray emission extends well beyond the
HCG galaxies \citep{PB93}.
The significant gas density at the group center is attested by the presence
of a cooling flow, as witnessed by the increase in surface brightness and the
cooler temperature of 
the inner  $50 \, h_{50}^{-1} \, \rm kpc$ \citep{PB93}.

The X-ray morphology of HCG~16 is very different from that of HCG~62.
The diffuse X-ray emission is situated within 
$\simeq 50 \, h_{50}^{-1} \, \rm kpc$
around the galaxies, plus in
one clump (C4) at $140 \, h_{50}^{-1} \, \rm kpc$ from the nearest galaxy.
This clumpy X-ray morphology strongly suggests that HCG~16 is far from
virialization. 


This argument against virialization supposes 
that intergalactic
gas, at the time of formation of the group, was able to settle in hydrostatic
equilibrium in the shallow potential of HCG~16. 
This point of view can be
challenged if the specific entropy of the intergalactic gas at this epoch is
high enough to prevent the gas from collapsing with the dark matter 
(Ponman, private communication). 
Indeed, assuming the infalling gas has the density of the
Universe at the epoch of group collapse, there is a maximum temperature above
which the specific entropy of the infalling gas will be higher than the
specific entropy of the gas settled in equilibrium in the potential of the 
group. Since the gas entropy cannot decrease (unless the gas radiates), this
constraint gives a maximum temperature above which the gas cannot settle in
the dark matter potential. 

This (relatively)
high specific entropy intergalactic gas still lacks direct observation, but
it is indirectly
inferred from the changes in the X-ray surface brightness profiles
from clusters to groups \citep*{PCN99}.
Also, the high temperature level
is required in the interpretation of a number of observational facts. For
example, the negative result of the  Gunn-Peterson effect implies that the
Universe had an overall reionization (and therefore probably reheating) phase
before $z \simeq 5$. Moreover, the steepening of the $L_X-T$ relation from
clusters to groups (PBEB) has so far only been explained in models where the
infalling gas was preheated at a temperature of $\simeq 0.5 \,{\rm keV}$
\citep{CMT97,CMT98,BBP99}. Thus, provided that HCG~16 is forming today, we can
understand its irregular X-ray morphology as a consequence of its low virial
temperature, which does not exclude that the underlying dark matter potential
is relaxed. 
Finally, hydrodynamical cosmological
simulations indicate that most of the intergalactic gas is in the
$10^5-10^7\,\rm K$ temperature range \citep{CO99}.

If intergalactic gas cannot collapse onto the group, then the
diffuse gas observed in the group originates primarily from the
galaxies 
of the group, in the form of shock-heated tidally stripped gas or hot
winds from collective supernovae ejecta. 
In the latter case,  the diffuse hot gas should be metal-rich.
Our spectral fits do not constrain the metallicity of the group, whereas
PBEB found $Z<0.17$ for the
diffuse gas
that they detected.
Since half of their detected emission arises from the radio sources 
in regions C1 and C5, we cannot confirm the low metallicity of the HCG~16
diffuse gas.
Better signal-to-noise observations are required to answer
this question, in particular better constraints on metallicity and its spatial
variation in the group. 

Whether or not the group is near virial equilibrium, \emph{the 
hot gas appears too clumpy to be itself in hydrostatic equilibrium within a
nearly spherical potential}.

\subsubsection{Chance alignment models}
\label{camod}

Difficulties in the understanding of the properties of HCGs (in particular
their short crossing-times) have led some authors to the conclusion that most
of these objects are chance alignments along the line-of-sight within larger
structures, namely loose groups \citep{M86}, clusters \citep{WM89} or
cosmological 
filaments \citep{HKW95}.

Loose groups are obviously not dense enough to be
globally near full collapse (although their cores may have already collapsed
and formed interacting binaries), and are more likely to be in their early
phases of collapse \citep{DRGF93,M93_Aussois_Dyn,M94_Moriond,M95_Chalonge},
as is our Local Group, or even 
in the late stages of 
their Hubble expansion before their turnaround \citep{VB86}. 
Thus, one does not expect to see diffuse gas in a
compact group caused by a chance 
alignment within a collapsing near-spherical loose group.
Indeed, if loose groups have diffuse intergalactic gas, this gas is too
tenuous to be observed. Moreover, this gas should remain fairly cold until it
relaxes with the group potential, and this re-heating
must await the virialization of
the group.
Therefore, the gas that one expects to observe in X rays within loose groups
will be 
associated with the dense substructures within these groups, {\it i.e.\/,}
galaxies and interacting binaries (which are expected in chance alignments,
\citealp{M92_DAEC}), and with gas ejected by supernovae 
near interacting galaxies ({\it e.g.\/,} \citealp{RP98,HC99}). 

The X-ray properties of chance alignments within cosmological filaments 
depend strongly on the dynamical and thermal state of the cosmological
filament. Such filaments appear clearly in cosmological simulations.
For example, the hydrodynamical simulations of \cite{CO99} show that most of
the intergalactic gas is not only at temperatures of $10^5-10^7\,\rm K$, but
also within filamentary structures.
\citeauthor{CO99} argue that this gas was shock-heated mainly by structure
collapse and possibly 
also by 
supernovae.

It presents a major challenge to detect gas, either
within filaments of loose groups, at temperatures well below $0.3\,\rm keV$,
since the gas in these systems should be too tenuous to be an efficient
emitter or an efficient
absorber. Moreover, its typical temperature is too cool to observe in X-rays,
and its 
detection is difficult in the EUV because of contamination from 
emission from the galactic corona and strong absorption from galactic HI.
%
\emph {Hence, one does not expect to detect widespread diffuse X-ray emission
in 
compact groups caused by chance alignments within loose groups or
cosmological filaments.}
Therefore, in these chance alignments scenarios,
the diffuse X-rays detected in HCG~16 are associated with interacting pairs
of galaxies, perhaps emitted by gas that was
stripped by tidal interactions or ejected
by galactic winds generated by supernova explosions. 

Our revised X-ray luminosity for HCG~16 brings the group closer to the
extrapolation of the cluster luminosity-temperature relation and to the
the group $L_X-T$ relation of \cite{MZ98}.
Moreover, if there is a universal luminosity-temperature relation spanning
the range from individual galaxies to binaries to real compact groups such as
HCG~62 to rich clusters, then one expects that a chance alignment of $N$
equal luminosity systems along the line-of-sight will produce a group that
will be located a factor $N$ in luminosity
above this universal
luminosity-temperature relation.
This is  consistent with the position of HCG~16 in
the $L_X$-$T$ diagram relative to the relation of \citeauthor{MZ98}, but
the X-ray luminosity of HCG~16 is 300 times too large for its low temperature
in comparison with PBEB's $L_X-T$ relation, and chance alignments cannot
explain such a large luminosity excess.

In Paper~II, we investigate in more detail the
possibility that HCG~16 occurs as a chance alignment within
a looser group or a
cosmological filament .

\subsection{Concluding remarks}

Due to their low virial temperature ($T \simeq 1 \, {\rm keV}$), groups of
galaxies and their X-ray emission are of extreme importance in the study of
processes affecting the baryonic diffuse content of systems of galaxies, such
as reheating and early energy injection
\citep{CMT97,CMT98,BBP99,PCN99}. 
Earlier analyses of the ROSAT {\sf PSPC} observation of Hickson Compact
Groups 
\citep{SC95,PBEB96} were based on counting photons in a circle surrounding
the optical center of the group.  
The present detailed spatial analysis of X-ray emission in HCG~16 highlights
the importance of a multi-wavelength study, in particular for rejecting
point sources identified in the optical and radio wavebands. 
The clumpy nature of the hot gas in HCG~16 appears quite different from the
more regular diffuse emission seen in other compact groups, and HCG~16 loses
its previous characteristic of being the sole spiral-only compact group with
hot gas tracing a fairly regular potential well 
(see \citealp{PBEB96,Mulchaey99}).
Our study raises nevertheless the question that other compact groups
previously detected in X-rays may be significantly contaminated by superimposed
X-ray (point or extended) sources. 

Another question raised by this study is the inclusion of HCG~16 in
the $L_X-T$ relation of groups and clusters. Indeed, this relation measures
the trend between dark matter total mass and mean baryonic density \emph{in
systems where the diffuse gas is in equilibrium within the dark matter
potential}. Since the diffuse hot gas of HCG~16 
appears to lie in several ($N$) clumps, one would then expect that the group
luminosity be a factor $N$ above the $L_X-T$ relation, for given temperature.
If PBEB's $L_X-T$ relation is correct, then the factor of 300 excess
luminosity for its temperature
in comparison with the extrapolation of the PBEB's trend is too large to be
consistent with the idea that most of the X-ray
emission originates from a few clumps of hot gas in equilibrium within an
underlying potential.
Moreover, in the frame of theories explaining the $L_X-T$ relation in groups
\citep{CMT97,CMT98,BBP99}, gas cannot settle in hydrostatic equilibrium
into a system with a virial
temperature smaller than 
$\simeq 0.5\,\rm keV$  (see Sect.~\ref{virgroup}). 
These two arguments naturally lead to the conclusion that the diffuse gas
observed in HCG~16 originates
mostly from the galaxies, either through tidal stripping or in galactic winds
driven by supernova explosions.


The launch of new X-ray satellites ({\sf Chandra})
with its high spatial and spectral resolution and XMM with its greatly enhanced
sensitivity and spectral resolution will eventually allow to answer these
questions. 
The precise nature 
of the diffuse emission seen in HCG~16 is
still difficult to constrain, because of the low signal-to-noise of the
ROSAT/{\sf PSPC} observations.
We expect to pursue our study of HCG~16 using archival ASCA data, as well as
a {\sf Chandra} observation, which we have obtained on this compact group.

\begin{acknowledgements}
This research constitutes part of the PhD thesis of S.D.S.
We acknowledge useful discussions with
Joel Bregman,
Malcolm Bremer,
Florence Durret,
Daniel Gerbal,
Mark Henriksen,
St\'ephane Leon,
Gast\~ao Lima Neto,
Claudia Mendes de Oliveira,
Vincent Pislar, and
Trevor Ponman. 
We are most grateful to Trevor Ponman for his excellent
refereeing, which led to substantial
improvement of our work.
Thanks also to
Michel Fioc for providing us digital output from his
{\sf PEGASE} spectro-photometric evolution model,
Gast\~ao Lima Neto for use of his graphics package,
Eric Slezak for his {\sf TRANSWAVE} \emph{\`a trous} wavelet package,
and
Barbara Williams for sending us her VLA
21cm maps and mass estimates in advance of
publication.
This research was supported in part by a grant from the French
GdR Cosmologie (awarded to G.A.M.).
We have made use of the {\sf NASA/IPAC Extragalactic Database (NED)}
which is 
operated by the Jet Propulsion Laboratory, California Institute of
Technology, under 
contract with the National Aeronautics and Space Administration, the
{\sf SIMBAD} database operated by the Centre de Donn\'ees Stellaires in
Strasbourg, France, the ROE/NRL {\sf COSMOS} UKST Southern Sky Object
Catalog,
the {\sf NRAO VLA Sky Survey (NVSS)}
and
NASA's {\sf Astrophysics Data System (ADS)} Abstract Service.
\end{acknowledgements}
\bibliography{master}

\begin{thebibliography}{62}
\expandafter\ifx\csname natexlab\endcsname\relax\def\natexlab#1{#1}\fi

\bibitem[{Arnaud} \& {Rothenflug}(1985)]{AR85}
{Arnaud}, M., {Rothenflug}, R., 1985, A\&AS 60, 425

\bibitem[{Bahcall} \& {Tremaine}(1981)]{BT81}
{Bahcall}, J.N., {Tremaine}, S., 1981, \apj 244, 805

\bibitem[{Bahcall} et~al.(1984){Bahcall}, {Harris}, \& {Rood}]{BHR84}
{Bahcall}, N.A., {Harris}, D.E., {Rood}, H.J., 1984, \apjl 284, L29

\bibitem[{Balogh} et~al.(1999){Balogh}, {Babul}, \& {Patton}]{BBP99}
{Balogh}, M., {Babul}, A., {Patton}, D., 1999, \mnras in press,
  astro-ph/9809159

\bibitem[{Balucinska-Church} \& {McCammon}(1992)]{BCM92}
{Balucinska-Church}, M., {McCammon}, D., 1992, \apj 400, 699

\bibitem[{Barnes}(1985)]{Barnes85}
{Barnes}, J., 1985, \mnras 215, 517

\bibitem[{Barnes}(1989)]{Barnes89}
{Barnes}, J.E., 1989, \nat 338, 123

\bibitem[{Bode} et~al.(1994){Bode}, {Berrington}, {Cohn}, \& {Lugger}]{BBCL94}
{Bode}, P.W., {Berrington}, R.C., {Cohn}, H.N., {Lugger}, P.M., 1994, \apj 433,
  479

\bibitem[{Carnevali} et~al.(1981){Carnevali}, {Cavaliere}, \&
  {Santangelo}]{CCS81}
{Carnevali}, P., {Cavaliere}, A., {Santangelo}, P., 1981, \apj 249, 449

\bibitem[{Cavaliere} et~al.(1997){Cavaliere}, {Menci}, \& {Tozzi}]{CMT97}
{Cavaliere}, A., {Menci}, N., {Tozzi}, P., 1997, \apjl 484, L21

\bibitem[{Cavaliere} et~al.(1998){Cavaliere}, {Menci}, \& {Tozzi}]{CMT98}
{Cavaliere}, A., {Menci}, N., {Tozzi}, P., 1998, \apj 501, 493

\bibitem[{Cen} \& {Ostriker}(1999)]{CO99}
{Cen}, R., {Ostriker}, J.P., 1999, \apj 514, 1

\bibitem[{Condon} et~al.(1998){Condon}, {Cotton}, {Greisen}, {Yin}, {Perley},
  {Taylor}, \& {Broderick}]{Condon+98}
{Condon}, J.J., {Cotton}, W.D., {Greisen}, E.W., {Yin}, Q.F., {Perley}, R.A.,
  {Taylor}, G.B., {Broderick}, J.J., 1998, \aj 115, 1693

\bibitem[{de Carvalho} et~al.(1994){de Carvalho}, {Ribeiro}, \& {Zepf}]{dCRZ94}
{de Carvalho}, R.R., {Ribeiro}, A.L.B., {Zepf}, S.E., 1994, \apjs 93, 47

\bibitem[{Diaferio} et~al.(1994){Diaferio}, {Geller}, \& {Ramella}]{DGR94}
{Diaferio}, A., {Geller}, M.J., {Ramella}, M., 1994, \aj 107, 868

\bibitem[{Diaferio} et~al.(1993){Diaferio}, {Ramella}, {Geller}, \&
  {Ferrari}]{DRGF93}
{Diaferio}, A., {Ramella}, M., {Geller}, M.J., {Ferrari}, A., 1993, \aj 105,
  2035

\bibitem[{Ebeling} et~al.(1994){Ebeling}, {Voges}, \& {B\"ohringer}]{EVB94}
{Ebeling}, H., {Voges}, W., {B\"ohringer}, H., 1994, \apj 436, 44

\bibitem[{Fioc} \& {Rocca-Volmerange}(1997)]{FRV97}
{Fioc}, M., {Rocca-Volmerange}, B., 1997, \aap 326, 950

\bibitem[{Governato} et~al.(1996){Governato}, {Tozzi}, \& {Cavaliere}]{GTC96}
{Governato}, F., {Tozzi}, P., {Cavaliere}, A., 1996, \apj 458, 18

\bibitem[{Heisler} et~al.(1985){Heisler}, {Tremaine}, \& {Bahcall}]{HTB85}
{Heisler}, J., {Tremaine}, S., {Bahcall}, J.N., 1985, \apj 298, 8

\bibitem[{Henriksen} \& {Cousineau}(1999)]{HC99}
{Henriksen}, M., {Cousineau}, S., 1999, \apj 511, 595

\bibitem[{Hernquist} et~al.(1995){Hernquist}, {Katz}, \& {Weinberg}]{HKW95}
{Hernquist}, L., {Katz}, N., {Weinberg}, D.H., 1995, \apj 442, 57

\bibitem[{Hickson}(1982)]{H82}
{Hickson}, P., 1982, \apj 255, 382

\bibitem[{Hickson}(1997)]{H97rev}
{Hickson}, P., 1997, \araa 35, 357

\bibitem[{Hickson} et~al.(1989){Hickson}, {Kindl}, \& {Auman}]{HKA89}
{Hickson}, P., {Kindl}, E., {Auman}, J.R., 1989, \apjs 70, 687

\bibitem[{Hickson} et~al.(1992){Hickson}, {Mendes de Oliveira}, {Huchra}, \&
  {Palumbo}]{HMHP92}
{Hickson}, P., {Mendes de Oliveira}, C., {Huchra}, J.P., {Palumbo}, G.G., 1992,
  \apj 399, 353

\bibitem[{Leon} et~al.(1998){Leon}, {Combes}, \& {Menon}]{LCM98}
{Leon}, S., {Combes}, F., {Menon}, T.K., 1998, \aap 330, 37

\bibitem[{Liedahl} et~al.(1995){Liedahl}, {Osterheld}, \& {Goldstein}]{LOG95}
{Liedahl}, D.A., {Osterheld}, A.L., {Goldstein}, W.H., 1995, \apjl 438, L115

\bibitem[{Mamon}(1986)]{M86}
{Mamon}, G.A., 1986, \apj 307, 426

\bibitem[{Mamon}(1987)]{M87}
{Mamon}, G.A., 1987, \apj 321, 622

\bibitem[{Mamon}(1992)]{M92_DAEC}
{Mamon}, G.A., 1992, in: {Mamon}, G.A., {Gerbal}, D. (eds.), 2nd DAEC mtg.,
  Distribution of Matter in the Universe, Obs. de Paris, Paris,  p. 51

\bibitem[{Mamon}(1993)]{M93_Aussois_Dyn}
{Mamon}, G.A., 1993, in: {Combes}, F., {Athanassoula}, E. (eds.), Gravitational
  Dynamics and the N-Body Problem, Obs. de Paris, Paris,  p. 188,
  astro-ph/9308032

\bibitem[{Mamon}(1994)]{M94_Moriond}
{Mamon}, G.A., 1994, in: {Durret}, F., {Mazure}, A., {White}, S.D.M., {Tr\^anh
  Thanh V\^an}, J. (eds.), 14th Moriond Astrophysics Mtg., Clusters of
  Galaxies, Fronti\`eres, Gif-sur-Yvette,  p. 291, astro-ph/9511101

\bibitem[{Mamon}(1995)]{M95_Chalonge}
{Mamon}, G.A., 1995, in: {de Vega}, H., {S\'anchez}, N. (eds.), 3rd Paris
  cosmology colloq.,  p. 95, astro-ph/9511101

\bibitem[{Mamon} \& {Dos Santos}(1999)]{MDS99}
{Mamon}, G.A., {Dos Santos}, S., 1999, \aap to be submitted (Paper II)

\bibitem[{Mewe} et~al.(1985){Mewe}, {Gronenschild}, \& {Van Den Oord}]{MGO85}
{Mewe}, R., {Gronenschild}, E.H.B.M., {Van Den Oord}, G.H.J., 1985, A\&AS 62,
  197

\bibitem[{Mewe} et~al.(1986){Mewe}, {Lemen}, \& {Van Den Oord}]{MLO86}
{Mewe}, R., {Lemen}, J.R., {Van Den Oord}, G.H.J., 1986, A\&AS 65, 511

\bibitem[{Motch} et~al.(1998){Motch}, {Guillout}, {Haberl}, {Krautter},
  {Pakull}, {Pietsch}, {Reinsch}, {Voges}, \& {Zickgraf}]{Motch+98}
{Motch}, C., {Guillout}, P., {Haberl}, F., {Krautter}, J., {Pakull}, M.W.,
  {Pietsch}, W., {Reinsch}, K., {Voges}, W., {Zickgraf}, F.J., 1998, \aaps 132,
  341

\bibitem[{Mulchaey}(1999)]{Mulchaey99}
{Mulchaey}, J.S., 1999, in: {Valtonen}, M.J., {Flynn}, C. (eds.), IAU Coll. No.
  174, Small Galaxy Groups, ASP, San Francisco, in press

\bibitem[{Mulchaey} et~al.(1996){Mulchaey}, {Davis}, {Mushotzky}, \&
  {Burstein}]{MDMB96}
{Mulchaey}, J.S., {Davis}, D.S., {Mushotzky}, R.F., {Burstein}, D., 1996, \apj
  456, 80

\bibitem[{Mulchaey} \& {Zabludoff}(1998)]{MZ98}
{Mulchaey}, J.S., {Zabludoff}, A.I., 1998, \apj 496, 73

\bibitem[{Pildis} et~al.(1995){Pildis}, {Bregman}, \& {Evrard}]{PBE95}
{Pildis}, R.A., {Bregman}, J.N., {Evrard}, A.E., 1995, \apj 443, 514

\bibitem[{Ponman} \& {Bertram}(1993)]{PB93}
{Ponman}, T.J., {Bertram}, D., 1993, \nat 363, 51

\bibitem[{Ponman} et~al.(1996){Ponman}, {Bourner}, {Ebeling}, \&
  {B\"ohringer}]{PBEB96}
{Ponman}, T.J., {Bourner}, P.D.J., {Ebeling}, H., {B\"ohringer}, H., 1996,
  \mnras 283, 690 (PBEB)

\bibitem[{Ponman} et~al.(1999){Ponman}, {Cannon}, \& {Navarro}]{PCN99}
{Ponman}, T.J., {Cannon}, D.B., {Navarro}, J.F., 1999, \nat 397, 135

\bibitem[{Rana} \& {Basu}(1992)]{RB92}
{Rana}, N.C., {Basu}, S., 1992, \aap 265, 499

\bibitem[{Read} \& {Ponman}(1998)]{RP98}
{Read}, A.M., {Ponman}, T.J., 1998, \mnras 297, 143

\bibitem[{Ribeiro} et~al.(1996){Ribeiro}, {de Carvalho}, {Coziol}, {Capelato},
  \& {Zepf}]{RdC3Z96}
{Ribeiro}, A.L.B., {de Carvalho}, R.R., {Coziol}, R., {Capelato}, H.V., {Zepf},
  S.E., 1996, \apjl 463, L5 (RdC3Z)

\bibitem[{Rose}(1977)]{Rose77}
{Rose}, J.A., 1977, \apj 211, 311

\bibitem[{Saracco} \& {Ciliegi}(1995)]{SC95}
{Saracco}, P., {Ciliegi}, P., 1995, \aap 301, 348

\bibitem[{Shensa}(1992)]{Shensa92}
{Shensa}, M.J., 1992, Proceedings IEEE 40, 2464

\bibitem[{Slezak} et~al.(1990){Slezak}, {Bijaoui}, \& {Mars}]{SBM90}
{Slezak}, E., {Bijaoui}, A., {Mars}, G., 1990, \aap 227, 301

\bibitem[{Slezak} et~al.(1993){Slezak}, {de Lapparent}, \& {Bijaoui}]{SdLB93}
{Slezak}, E., {de Lapparent}, V., {Bijaoui}, A., 1993, \apj 409, 517

\bibitem[{Snowden} et~al.(1994){Snowden}, {McCammon}, {Burrows}, \&
  {Mendehall}]{SMBM94}
{Snowden}, S.L., {McCammon}, D., {Burrows}, D.N., {Mendehall}, J.A., 1994, \apj
  424, 714

\bibitem[{Starck} \& {Murtagh}(1994)]{SM94}
{Starck}, J.L., {Murtagh}, F., 1994, \aap 288, 342

\bibitem[{Starck} \& {Pierre}(1998)]{SP98}
{Starck}, J.L., {Pierre}, M., 1998, A\&AS 128, 397

\bibitem[{Stark} et~al.(1992){Stark}, {Gammie}, {Wilson}, {Bally}, {Linke},
  {Heiles}, \& {Hurwitz}]{Stark92}
{Stark}, A.A., {Gammie}, C.F., {Wilson}, R.W., {Bally}, J., {Linke}, R.A.,
  {Heiles}, C., {Hurwitz}, M., 1992, \apjs 79, 77

\bibitem[{Tully} \& {Fisher}(1977)]{TF77}
{Tully}, R.B., {Fisher}, J.R., 1977, \aap 54, 661

\bibitem[{Valtonen} \& {Byrd}(1986)]{VB86}
{Valtonen}, M.J., {Byrd}, G.G., 1986, \apj 303, 523

\bibitem[{Verdes-Montenegro}(1999)]{VM99}
{Verdes-Montenegro}, L., 1999, in: {Valtonen}, M.J., {Flynn}, C.(eds.), IAU
  Coll. No. 174, Small Galaxy Groups, ASP, San Francisco, in press

\bibitem[{Walke} \& {Mamon}(1989)]{WM89}
{Walke}, D.G., {Mamon}, G.A., 1989, \aap 225, 291

\bibitem[{Williams}(1998)]{Williams98}
{Williams}, B.A., 1998, private communication

\end{thebibliography}
\end{document}